\documentclass[twocolumn]{aastex63}
\usepackage{hyperref}
\usepackage{longtable}

\newcommand{\apg}  	{\ga}
\newcommand{\apll}  	{\la}
\newcommand{\eg}{{\it e.g.}}
\newcommand{\ie}{{\it i.e.}}
\newcommand{\no}{\nodata}
\newcommand{\Msol}{\hbox{M$_\odot$}}
\shorttitle{El Gordo Lens Model and a Background Galaxy Overdensity}
\shortauthors{Frye et al.}
\graphicspath{{./}{figures/}}

\begin{document}

\title{The JWST PEARLS View of the El Gordo Galaxy Cluster and of the Structure It Magnifies}

\author[0000-0003-1625-8009]{Brenda L.~Frye}
\affiliation{Department of Astronomy/Steward Observatory, University of Arizona, 933 N. Cherry Avenue, Tucson, AZ 85721, USA}

\author[0000-0002-2282-8795]{Massimo Pascale}
\affiliation{Department of Astronomy, University of California, 501 Campbell Hall \#3411, Berkeley, CA 94720, USA}

\author[0000-0002-7460-8460]{Nicholas Foo}
\affiliation{Department of Astronomy/Steward Observatory, University of Arizona, 933 N. Cherry Avenue, Tucson, AZ 85721, USA}

\author{Reagen Leimbach}
\affiliation{Department of Astronomy/Steward Observatory, University of Arizona, 933 N. Cherry Avenue, Tucson, AZ 85721, USA}

\author{Nikhil Garuda}
\affiliation{Department of Astronomy/Steward Observatory, University of Arizona, 933 N. Cherry Avenue, Tucson, AZ 85721, USA}

\author{Paulina Soto Robles}
\affiliation{Department of Astronomy/Steward Observatory, University of Arizona, 933 N. Cherry Avenue, Tucson, AZ 85721, USA}

\author{Jake Summers} 
\affiliation{School of Earth \& Space Exploration, Arizona State University, Tempe, AZ 85287-1404, USA}

\author{Carlos Diaz}
\affiliation{Department of Astronomy, University of Massachusetts, Amherst, MA 01003, USA}

\author[0000-0001-9394-6732]{Patrick Kamieneski}
\affiliation{School of Earth \& Space Exploration, Arizona State University, Tempe, AZ 85287-1404, USA}

\author[0000-0001-6278-032X]{Lukas J. Furtak}
\affiliation{Physics Department, Ben-Gurion University of the Negev, P. O. Box 653, Be’er-Sheva, 8410501, Israel}

\author[0000-0003-3329-1337]{Seth H.~Cohen}
\affiliation{School of Earth \& Space Exploration, Arizona State University, Tempe, AZ 85287-1404, USA}

\author{Jose Diego}
\affiliation{FCA, Instituto de Fisica de Cantabria (UC-CSIC), Av.  de Los Castros s/n, E-39005 Santander, Spain}

\author{Benjamin Beauchesne}
\affiliation{Institute of Physics, Laboratory of Astrophysics, Ecole Polytechnique Fed'erale de Lausanne (EPFL), Observatoire de Sauverny, CH-1290 Versoix, Switzerland}
\affiliation{Univ Lyon, Univ Lyon1, Ens de Lyon, CNRS, Centre de Recherche Astrophysique de Lyon UMR5574, F-69230 Saint-Genis-Laval, France}

\author[0000-0001-8156-6281]{Rogier A.~Windhorst} 
\affiliation{School of Earth \& Space Exploration, Arizona State University, Tempe, AZ 85287-1404, USA}

\author[0000-0002-9895-5758]{S. P. Willner}
\affiliation{Center for Astrophysics \textbar\ Harvard \& Smithsonian, 60 Garden Street, Cambridge, MA 02138, USA}

\author[0000-0002-6610-2048]{Anton M.~Koekemoer} 
\affiliation{Space Telescope Science Institute,
3700 San Martin Drive, Baltimore, MD 21218, USA}

\author{Adi Zitrin}
\affiliation{Physics Department, Ben-Gurion University of the Negev, P. O. Box 653, Be’er-Sheva, 8410501, Israel}

\author{Gabriel Caminha}
\affiliation{Max-Planck-Institut für Astrophysik, Karl-Schwarzschild-Str. 1, D-85748 Garching, Germany}

\author{Karina I.~Caputi}
\affiliation{Kapteyn Astronomical Institute, University of Groningen, Postbus 800, 9700 AV Groningen, The Netherlands}
\affiliation{The Cosmic Dawn Center, Niels Bohr Institute, University of Copenhagen, Julian Maries Vej 30, DK-2100 Copenhagen $\emptyset$, Denmark}

\author{Dan Coe}
\affiliation{Space Telescope Science Institute,
3700 San Martin Drive, Baltimore, MD 21218, USA}

\author[0000-0003-1949-7638]{Christopher J.~Conselice} 
\affiliation{Jodrell Bank Centre for Astrophysics, University of Manchester, Oxford Road, Manchester, M13\,9PL, U.K.} 

\author[0000-0003-2091-8946]{Liang Dai}
\affiliation{Department of Physics, University of California, 366 Physics North MC 7300, Berkeley, CA. 94720, USA}

\author{Herv\'{e} Dole}
\affiliation{Universit\'e Paris-Saclay, CNRS, Institut d'Astrophysique Spatiale, 91405, Orsay, France}

\author[0000-0001-9491-7327]{Simon P. Driver} 
\affiliation{International Centre for Radio Astronomy Research (ICRAR) and the
International Space Centre (ISC), The University of Western Australia, M468,
35 Stirling Highway, Crawley, WA 6009, Australia}

\author[0000-0001-9440-8872]{Norman A. Grogin} 
\affiliation{Space Telescope Science Institute,
3700 San Martin Drive, Baltimore, MD 21218, USA}

\author{Kevin Harrington}
\affiliation{European Southern Observatory, Alonso de C\'ordova 3107, Vitacura, Casilla 19001, Santiago de Chile, Chile}

\author[0000-0003-1268-5230]{Rolf A.~Jansen} 
\affiliation{School of Earth \& Space Exploration, Arizona State University, Tempe, AZ 85287-1404, USA}

\author{Jean-Paul Kneib}
\affiliation{Institute of Physics, Laboratory of Astrophysics, Ecole Polytechnique Fed'erale de Lausanne (EPFL), Observatoire de Sauverny, CH-1290 Versoix, Switzerland}
\affiliation{Aix Marseille Universite, CNRS, LAM (Laboratoire d’Astrophysique de Marseille) UMR 7326, F-13388 Marseille, France}

\author{Matt Lehnert}
\affiliation{Univ Lyon, Univ Lyon1, Ens de Lyon, CNRS, Centre de Recherche Astrophysique de Lyon UMR5574, F-69230 Saint-Genis-Laval, France}

\author{James Lowenthal}
\affiliation{Smith College, Northampton, MA 01063, USA}

\author[0000-0001-6434-7845]{Madeline A. Marshall} 
\affiliation{National Research Council of Canada, Herzberg Astronomy \&
Astrophysics Research Centre, 5071 West Saanich Road, Victoria, BC V9E 2E7,
Canada}
\affiliation{ARC Centre of Excellence for All Sky Astrophysics in 3 Dimensions
(ASTRO 3D), Australia}

\author {Felipe Menanteau}
\affiliation{Department of Astronomy, University of Illinois at Urbana Champaign, 1002 W. Green Street, Urbana, IL 61801, USA}
\affiliation{Center for Astrophysical Surveys, National Center for Supercomputing Applications, 1205 West Clark St., Urbana, IL 61801, USA}

\author {Bel\'{e}n Alcalde Pampliega}
\affiliation{European Southern Observatory, Alonso de C\'ordova 3107, Vitacura, Casilla 19001, Santiago de Chile, Chile}

\author[0000-0003-3382-5941]{Nor Pirzkal} 
\affiliation{Space Telescope Science Institute,
3700 San Martin Drive, Baltimore, MD 21218, USA}

\author{Mari Polletta}
\affiliation{INAF – Istituto di Astrofisica Spaziale e Fisica cosmica (IASF) Milano, Via A. Corti 12, 20133 Milan, Italy}

\author{Johan Richard}
\affiliation{Univ Lyon, Univ Lyon1, Ens de Lyon, CNRS, Centre de Recherche Astrophysique de Lyon UMR5574, F-69230 Saint-Genis-Laval, France}

\author[0000-0003-0429-3579]{Aaron Robotham} 
\affiliation{International Centre for Radio Astronomy Research (ICRAR) and the
International Space Centre (ISC), The University of Western Australia, M468,
35 Stirling Highway, Crawley, WA 6009, Australia}

\author[0000-0003-0894-1588]{Russell E. Ryan, Jr.} 
\affiliation{Space Telescope Science Institute,
3700 San Martin Drive, Baltimore, MD 21218, USA}

\author[0000-0001-7016-5220]{Michael J. Rutkowski} 
\affiliation{Minnesota State University-Mankato,  Telescope Science Institute,
TN141, Mankato MN 56001, USA}

\author{Christ\'{o}bal Sif\'{o}n}
\affiliation{Instituto de F\'{i}sica, Pontificia Universidad Cat\'{o}lica de Valpara\'{i}so, Casilla 4059, Valpara\'{i}so, Chile}

\author[0000-0001-9052-9837]{Scott Tompkins} 
\affiliation{School of Earth \& Space Exploration, Arizona State University, Tempe, AZ 85287-1404, USA}

\author{Daniel Wang}
\affiliation{Department of Astronomy, University of Massachusetts at Amherst, Amherst, MA 01003, USA}

\author[0000-0001-7592-7714]{Haojing Yan}
\affiliation{Department of Physics and Astronomy, University of Missouri, Columbia, MO 65211}

\author{Min S.~Yun}
\affiliation{Department of Astronomy, University of Massachusetts at Amherst, Amherst, MA 01003, USA}

\begin{abstract}

The massive galaxy cluster El Gordo ($z=0.87$) imprints multitudes of gravitationally lensed arcs onto James Webb Space Telescope (JWST) Near-Infrared Camera (NIRCam) images. Eight bands of NIRCam imaging were obtained in the ``Prime Extragalactic Areas for Reionization and Lensing Science'' (``PEARLS'') program.  PSF-matched photometry across Hubble Space Telescope (HST) and NIRCam filters supplies new photometric redshifts. A new light-traces-mass lens model based on 56 image multiplicities identifies the two mass peaks and yields a mass estimate within 500~kpc of $(7.0 \pm 0.30) \times 10^{14}$~\Msol. A search for substructure in the 140 cluster members with spectroscopic redshifts confirms the two main mass components. The southeastern mass peak that contains the BCG is more tightly bound than the northwestern one. The virial mass within 1.7~Mpc is (5.1\,$\pm$\,0.60)$\times$10$^{14}$~\Msol, lower than the lensing mass.  A significant transverse velocity component could mean the virial mass is underestimated. 
We contribute one new member to the previously known $z=4.32$ galaxy group. Intrinsic (delensed) positions of the five secure group members span a physical extent of $\sim$60~kpc. 
Thirteen additional candidates selected by spectroscopic/photometric constraints are small and faint with a mean intrinsic luminosity $\sim$2.2 mag fainter than $L^*$. NIRCam imaging admits a fairly wide range of brightnesses and morphologies for the group members, suggesting a more diverse galaxy population in this galaxy overdensity.
 
\end{abstract}


\section{Introduction} \label{sec:intro}

The $z=0.87$ galaxy cluster ACT-CL J0102$-$4915, known as El Gordo, was discovered by its large  Sunyaev--Z'eldovich effect (SZE) decrement observed by the Planck mission \citep{Marriage2011}.  The cluster has the  highest X-ray luminosity of any cluster at $z> 0.6$ \citep{Menanteau2012}, and filaments of synchrotron radio emission  flank the two major cluster components. Such filaments are typically a consequence of cluster-scale dynamical disturbances \citep{Linder2014}. Mass estimates by  independent methods, including applying the virial theorem to the velocity dispersion and strong- and weak-lensing approaches, place El Gordo at a mass ${\apg}10^{15}$~\Msol, close to exceeding the maximum mass allowable by the $\Lambda$CDM model at El Gordo's redshift \citep[\eg,][]{Mortonson2011}.  \citet{Diego2023} gives a complete summary of the mass estimates of El Gordo and the associated caveats on such measurements.

The cluster's galaxy distribution consists of  prominent  southeastern (SE) and northwestern (NW) components, but there is some debate over which is dominant.  Weak-lensing and dynamical mass estimates have historically favored the NW component with a mass ratio of SE:NW close to 0.6:1 \citep{Menanteau2012, Jee2014}, while more recent weak- and strong-lensing studies promote the SE component  to a ratio closer to 1:1 \citep{Zitrin2013, Cerny2018,Diego2020,Kim2021,Diego2023}. \citet{Diego2023} pointed out that both values may be true if the SE component is more massive on smaller physical scales but gradually loses out to the NW one with increasing radius. The dynamical stage of this binary cluster is also under debate. While there is consensus of a major merger, it is not yet settled whether the collision was head-on or off-axis \citep[][and references therein]{Zhang2018}.  The cluster appears to be post-first-passage \citep{Molnar2015} and post-maximum-separation and is now on its return phase \citep{Ng2015}, although the unusual X-ray morphology and positions of the radio relics prevent strong constraints on its evolutionary stage \citep{Kim2021}.

\begin{figure}[h]
\centering\includegraphics[scale =0.31]{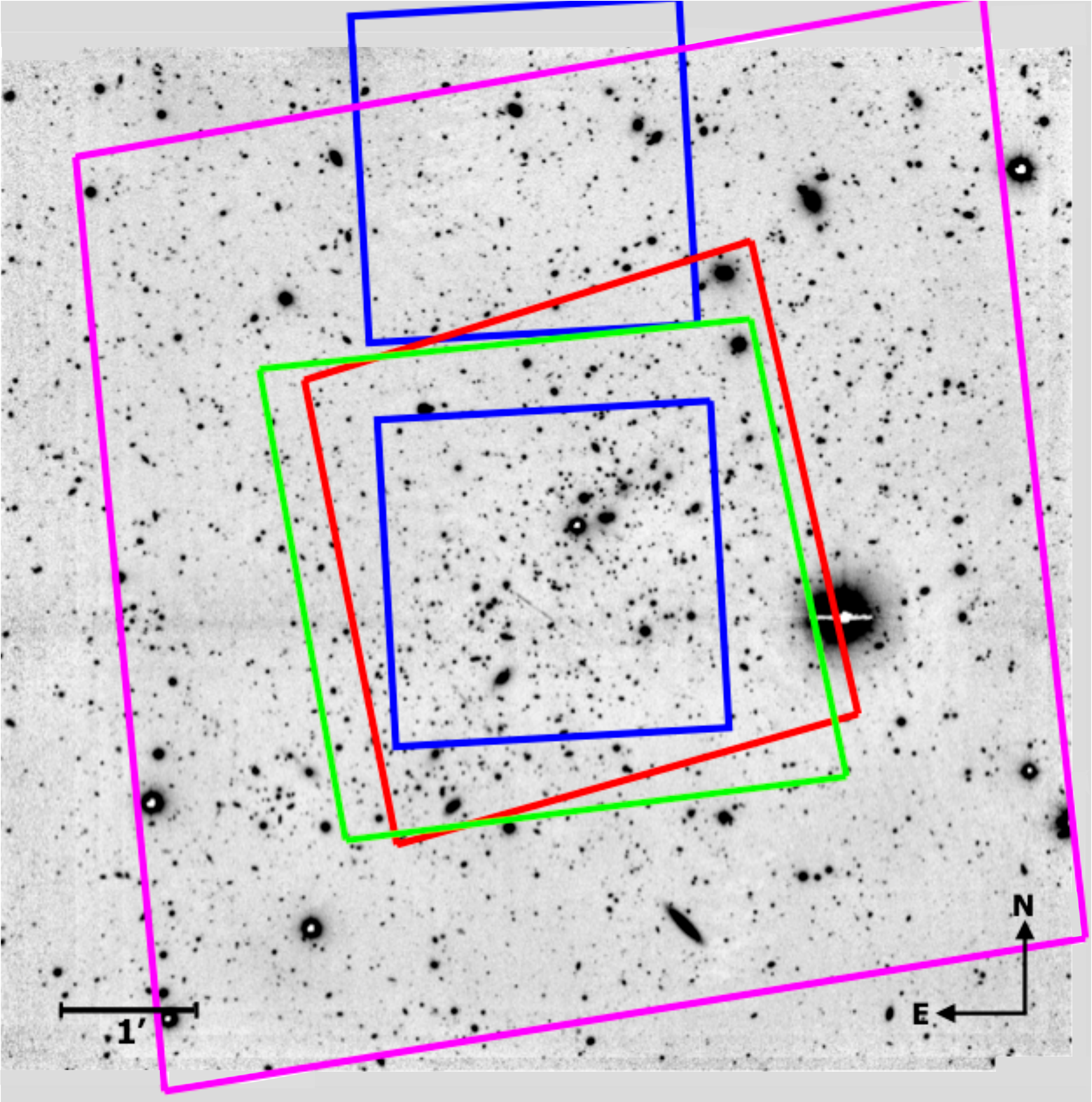}
\caption{ 
HST and JWST/NIRCam coverage of the field.  The background negative image is an $r$-band image from the Dark Energy Survey (DES) that has a field-of-view of 3.6 Mpc $\times$ 3.4 Mpc at the cluster redshift. The pair of blue squares depicts the two NIRCam fields of view, and other squares depict the HST imaging:  F435W (green), F814W (red), and F606W (magenta) in order of increasing field coverage.}
\label{fig_foot}
\end{figure}

The first estimate of El Gordo's mass distribution by strong-lensing modeling  was based on nine cases of a single galaxy appearing in multiple locations, called an ``image system," and was constructed prior to the availability of spectroscopic redshifts of the image systems to anchor the model \citep{Zitrin2013}. Nonetheless, this initial lens model confirmed the elongation of the cluster in the direction of the ongoing merger and the high mass. \citet{Cerny2018} extended the number of image systems by a factor of two, and \citet{Diego2020} contributed additional lensing constraints and confirmed the high mass.  Even with these steps forward, the lack of spectroscopic redshifts of the image systems limited the accuracy of the resulting lens models and their ability to recover the lensed image positions \citep{Johnson2016}.

An advance was made by \citet{Caminha2022}, who measured spectroscopic redshifts for 23 image systems in El Gordo using the Very Large Telescope (VLT) Multi-Unit Spectroscopic Explorer (MUSE) instrument. This result was made feasible by the integral field unit spectroscopic approach that obtains spectra for all objects in the field-of-view. Four of the systems are at a similar redshift of $z=4.32$, uncovering a strongly-lensed grouping of galaxies behind the cluster  \citep{Caputi2021}.  
By fitting spectral energy distributions (SEDs) to the Hubble Space Telescope (HST) photometry, \citet{Caputi2021} found the galaxies to be relatively low in mass (${\sim}10^7$--$10^{10}$~M$_{\odot}$) and to have star formation rates (SFRs) of 0.4--24 M$_{\odot}$~yr$^{-1}$ that qualify two of them as starbursts. This result led \citeauthor{Caputi2021} to suggest that these four galaxies may be experiencing enhanced star formation as a result of galaxy--galaxy interactions. The typical morphology of the galaxies---a  blue, compact clump superimposed on a more extended and redder component---supports this claim, but the sources are too red and faint for the HST data to separate the components. High-resolution imaging that extends  redward of the 4000~\AA\ and Balmer breaks ($\lambda>2$~\micron\ observed) is needed to characterize the stellar populations and morphologies of these galaxies.

The first lensing analysis of El Gordo to incorporate James Webb Space Telescope (JWST) Near Infrared Camera \citep[NIRCam; {\eg,}][]{Rieke05} imaging \citep{Diego2023} was completed using the non-parametric WSLAP+ approach. The authors constructed a base lens model by incorporating the 23 image systems of \citet{Caminha2022} and identified an additional 37 new image systems. Two lens models using all 60 systems assign roughly equal mass to the NW and SE components and confirm El Gordo's high mass. \citeauthor{Diego2023} also reported photometric redshift estimates using only the HST bands and  separately using only NIRCam bands. 

El Gordo is one of seven galaxy clusters  selected by the Prime Extragalactic Areas for Reionization and Lensing Science (PEARLS) project \citep{Windhorst2023} on account of its significant lensing strength and large critical-curve perimeter. This paper presents new PSF-matched photometry for El Gordo across HST and NIRCam filters, enabling more reliable photometric redshifts and a search for additional members of the $z=4.3$ galaxy overdensity.  An independent lensing analysis is carried out using a ``light traces mass'' (LTM) lens model instead of WSLAP+, and a new spectroscopic analysis is used to search for substructures. This paper is organized as follows. \S\ref{sec:obs} introduces the NIRCam data and the relevant ancillary imaging data. \S\ref{sec:phot} describes the new NIRCam photometry and PSF-matched photometry that incorporates the bluer HST filters and the generation of the object catalog. The construction of the LTM lens model appears in \S\ref{sec:LTM}. The spectroscopic analysis is given in \S\ref{sec:spec}, the new constraints on the cluster physical properties are discussed in \S\ref{sec:props}, and the search for new members of the galaxy overdensity at $z=4.3$ appears in \S\ref{sec:odense}. \S\ref{sec:finis} summarizes the results. This paper uses the AB magnitude system throughout, and we assume a  flat $\Lambda$CDM cosmology with $H_0=67$~km~s$^{-1}$\,Mpc$^{-1}$,  $\Omega_{m,0}=0.32$, and $\Omega_{\Lambda,0}=0.68$ \citep{PlanckCollaboration2018}.

\section{Observations and Reductions} \label{sec:obs}
\subsection{NIRCam} \label{sec:NIRCam}
NIRCam observations were obtained on 2022 July 29 as a part of the JWST PEARLS program (PI: Windhorst, Program ID 1176). The observing window was selected by STScI in order to reduce stray light expected from a nearby bright star in projection. Exposures were taken in F090W, F115W, F150W, F200W in the short wavelength (SW) module and F277W, F356W, F444W, and F410M in the long wavelength (LW) module. Point-source limiting magnitudes are filter dependent but approximately $m_{\rm AB}=28.0$--28.9 mag.  

\begin{figure*}[h]
\centering
\includegraphics[viewport=165 5 1000 836, scale=0.53]{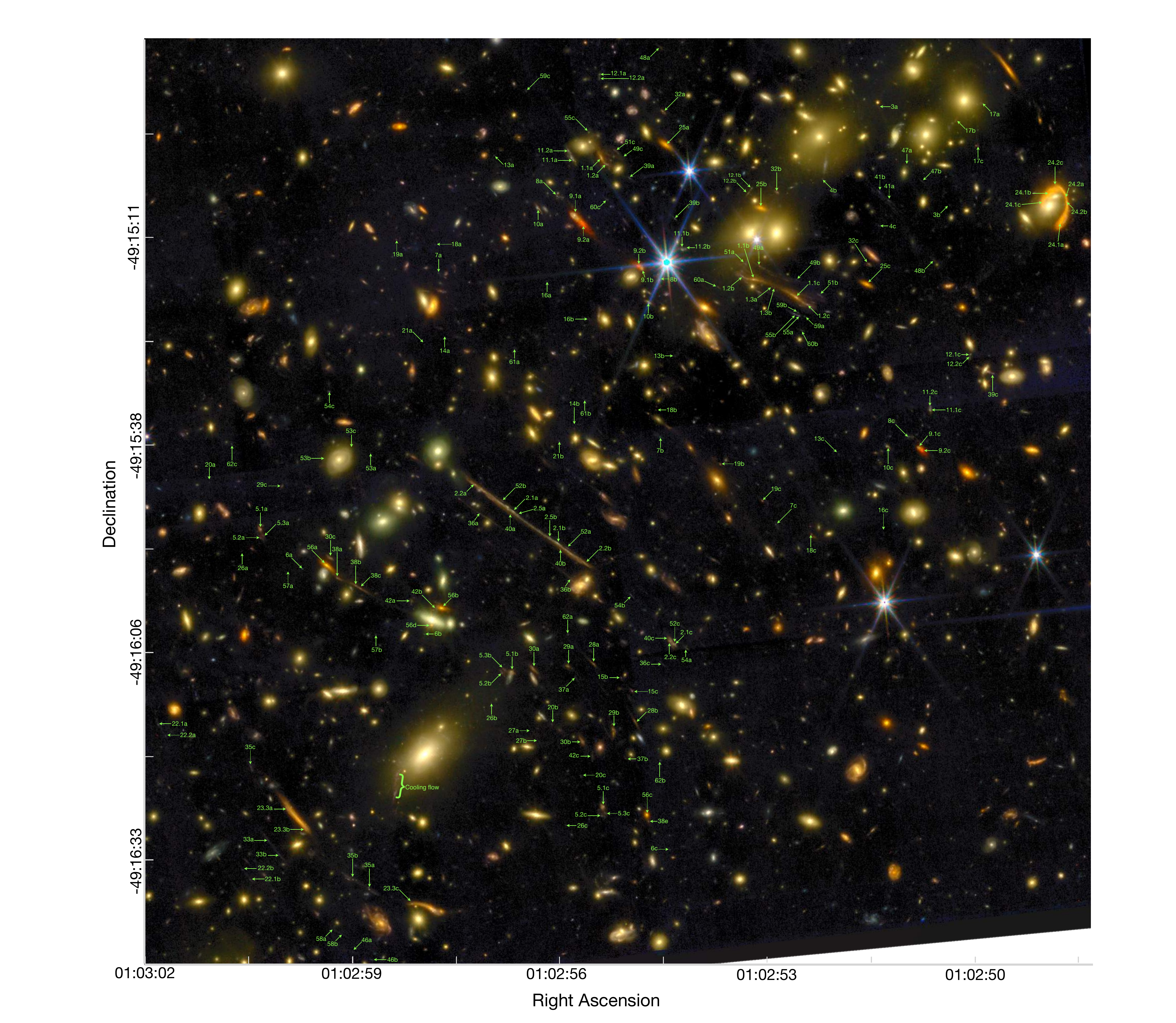}
\caption{NIRCam color image of the central region of El Gordo. The image is 2.2\,$\times$\,2.2 arcmin on a side, and the orientation is North up, East to the left.  Colors follow the prescription in Trilogy \citep{Coe2012} with  red showing F410M+F444W, green showing F200W+F277W+F356W, and blue showing F090W+F115W. All colors are multiplied by the sum of all the wide NIRCam filters (F090W + F115W + F150W + F200W + F277W + F356W + F444W) to the 1/4 power.  The 56 image systems used in our lens model are labeled in green, as is the cooling flow detected in HST imaging. The axes give the right ascension and declination in J2000 coordinates.}
\label{fig_map}
\end{figure*}

\begin{figure}[h]
\centering\includegraphics[scale=0.39]{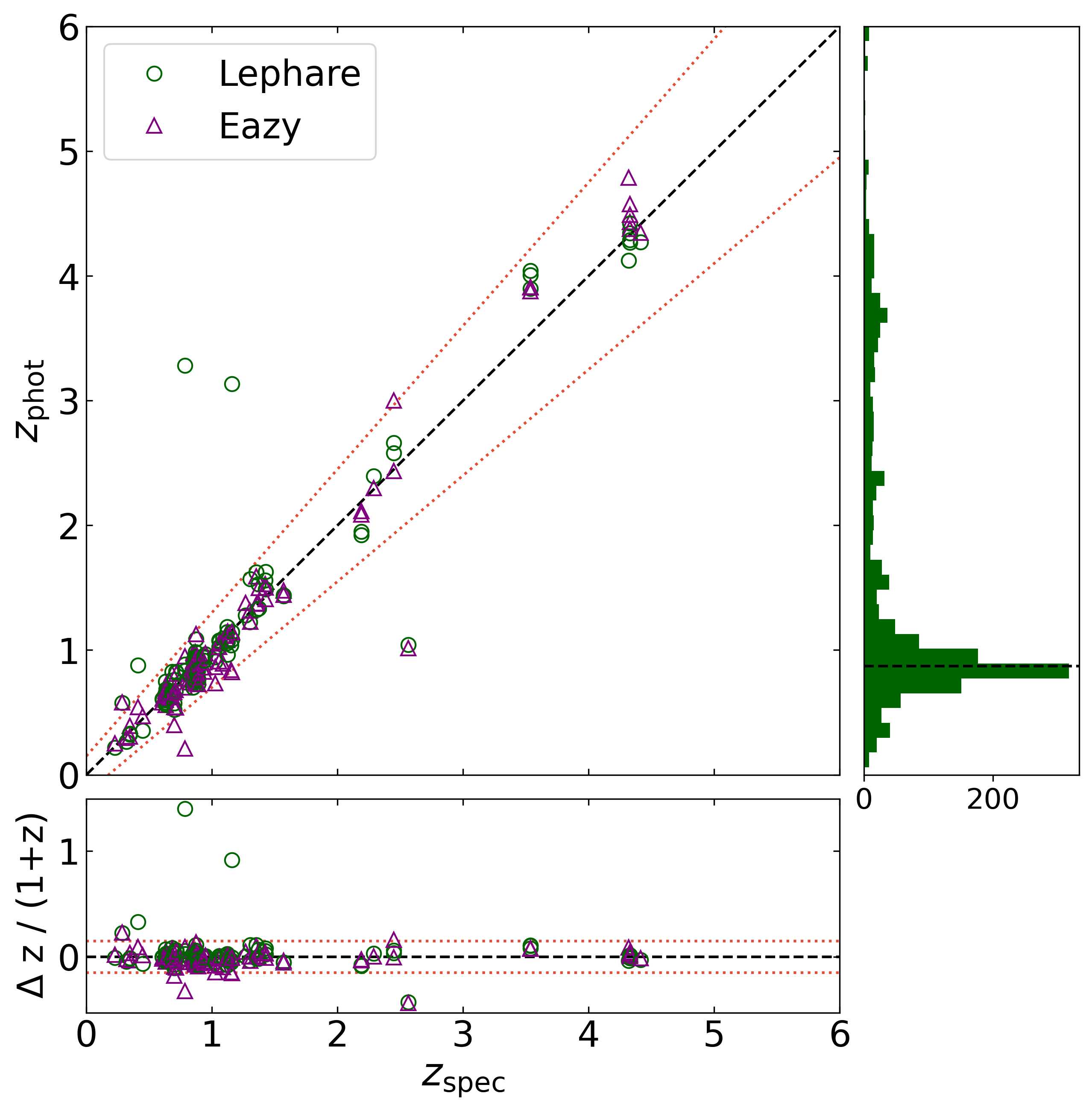}
\caption{Photometric redshifts vs.~spectroscopic redshifts. Points depict redshifts as indicated in the legend.  The dashed line shows equality, and the dotted lines delineate the region where $\vert\delta z \vert/(1+z)< 0.15$, the goodness-of-fit criterion used by  \citet{Pascale2022a}. Bottom panel shows $\vert\delta z \vert/(1+z)$ directly.  The panel on the right \edit1{gives} the histogram of photometric redshifts, which peaks at the El Gordo cluster redshift (shown by the dashed line).
}
\label{fig_photoz}
\end{figure}

The images were reduced by our team as discussed in detail by \citet{Windhorst2023}.  Briefly, the data were retrieved from the {Mikulski} Archive for Space Telescopes (MAST), corrected for $1/f$ noise by first using the prescription of C.\ Willott\footnote{\url{https://github.com/chriswillott/jwst.git}}.  We then ran the ProFound code, which makes a second round of corrections of any remaining residuals in the relevant rows and columns and also flattens the background and corrects for detector-level offsets, wisps, and snowballs \citep{Robotham2017,Robotham2018}. Finally, the frames were astrometrically aligned onto a common astrometric reference frame and drizzled into combined mosaics, following similar methodology to that first described by \cite{Koekemoer2011}, updated to use the JWST pipeline\footnote{\url{https://github.com/spacetelescope/jwst}}. Figure~\ref{fig_foot} depicts the field coverage against the backdrop of the $r$-band Dark Energy Survey (DES) image.  The single, central NIRCam pointing covers both the SE and NW cluster components and the many prominent arcs. Figure~\ref{fig_map} shows the NIRCam color image. Some of the data presented in this paper were obtained from the Mikulski Archive for Space Telescopes (MAST) at the Space Telescope Science Institute. The specific observations analyzed can be accessed via \dataset[10.17909/x49n-d207]{https://doi.org/10.17909/x49n-d207}.

\subsection{Ancillary Data} \label{sec:Ancillary}

We augmented the NIRCam data with bluer HST Wide Field Camera (WFC) F435W, F606W, and F814W imaging drawn from the literature (PID 14096; PID 12477). Data in other ACS-WFC, WFC3-IR and Spitzer/IRAC filters also exist which substantially overlap with the data used here, but they are superseded by the NIRCam imaging and so not included here.   Spectroscopy is available from \citet{Sifon2013} and \citet{Caminha2022}. 

Three Chandra observations were acquired (ObsID: 12258, 14022, 14023: PI: J.\ Hughes) with a total of 351~ks of combined exposure time. We reprocessed the observations to generate level-2 event files using the \texttt{chandra\_repro} script available in the Chandra Interactive Analysis of Observations \citep[CIAO;][]{Fruscione2006} software version 4.14 with CALDB v.~4.9.8. The three observations were merged using the \texttt{merge\_obs} script to produce a broad-band (0.5--7.0~keV), co-added, clipped counts image and a corresponding exposure map. We detected X-ray point sources using the \texttt{wavdetect} script and visually inspected the source list to create a list of point sources that did not pertain to the cluster. We removed and replaced the pixels of these sources with interpolated values from the surrounding background regions of each source using the \texttt{dmfilth} script. The resulting image was divided by the corresponding exposure map to produce an exposure-corrected image and smoothed using \texttt{aconvolve} with a Gaussian kernel. 

\section{Photometry} \label{sec:phot}
\subsection{NIRCam Photometry} \label{sec:phot_NIRCam}
Photometry is delivered individually for each NIRCam filter as a part of the JWST pipeline image reductions. Using only the eight NIRCam bands tends to underestimate the redshifts relative to the spectroscopic values \citep{Diego2023}, thereby motivating photometry extending to the bluer HST filters. These are the HST ACS filters F435W, F606W, and F814W, which  provide coverage blueward of the 4000~\AA\ and Balmer breaks at the redshift of the background galaxy overdensity at $z=4.32$. Deriving accurate photometric redshifts requires consistent photometry in all filters.

The JWST NIRCam F200W image was assigned as the reference image. In this filter, the imaging performance  is diffraction limited, defined as having a Strehl ratio $>$0.80. To find sources, SExtractor \citep{Bertin1996}  was used in a two-step HOT+COLD process following the prescription of \citet{Galametz2013}.  The object catalog reports AUTO magnitudes, which are measured in a Kron-like elliptical aperture.

For other JWST filters, each image was convolved with a kernel to match the PSF of the F200W image. The kernels were derived  in Fourier space by applying the convolution theorem \citep{Pascale2022a}.  The original rms maps remain adequate because all NIRCam filters have similar PSF FWHMs.  Isophotal magnitudes were derived from forced aperture photometry in SExtractor's dual-image mode using the F200W image and its associated weight map for detection but each other NIRCam image and its weight map for the measurements.  The ISO colors were then added to the F200W AUTO magnitudes to derive magnitudes in each filter.
This was done because isophotal magnitudes generally yield more accurate colors, while AUTO magnitudes tend to better estimate total flux \citep{Coe2006}. Magnitude uncertainties were computed by SExtractor based on the rms maps. Systematics are broadly accounted for in the photometric redshift estimation that follows by enforcing a  minimum uncertainty of 0.05~mag for each photometric measurement.



\subsection{HST Photometry} \label{sec:phot_HST}

To prepare the HST imaging for multiband photometry, the HST and NIRCam imaging were both aligned to the GAIA reference frame. The PSFs were then generated in all filters.  We first attempted to measure the PSFs in the HST images from the data by co-adding the profiles of isolated and unsaturated stars.  The search for suitable stars was done with the photutils \texttt{}{DAOStarFinder} \citep{Bradley2022}. This exercise yielded fewer than five stars in each filter, too few to give accurate PSFs.  Instead we adopted the model PSFs provided by STScI, which provide better representations of the point-source profile.\footnote{Empirical models for the WFC3/UVIS and WFC/ACS PSFs \citep{Krist2011}.} We also attempted to measure the  NIRCam PSFs from the data and again found too few stars, prompting us to model the PSFs using the \texttt{WebbPSF} software. 

The photometry in each of the HST filters proceeded as for the NIRCam photometry (\S3.1), measuring ISO color differences with respect to the F200W reference image.  This avoids measuring fluxes directly from the HST images, on which  fainter objects can get lost in the noise. 
This procedure assumes that the detection image is a good model for the HST images.
The coverage in the bluest available band, F435W, is relatively shallow, and none of the members of the galaxy overdensity at $z = 4.32$ was detected in this filter. 
Obtaining deeper F435W imaging would improve the SED fits discussed below, especially for the $z\ge4$  sources. 

\begin{figure}[tb]
\centering\includegraphics[scale=0.56]{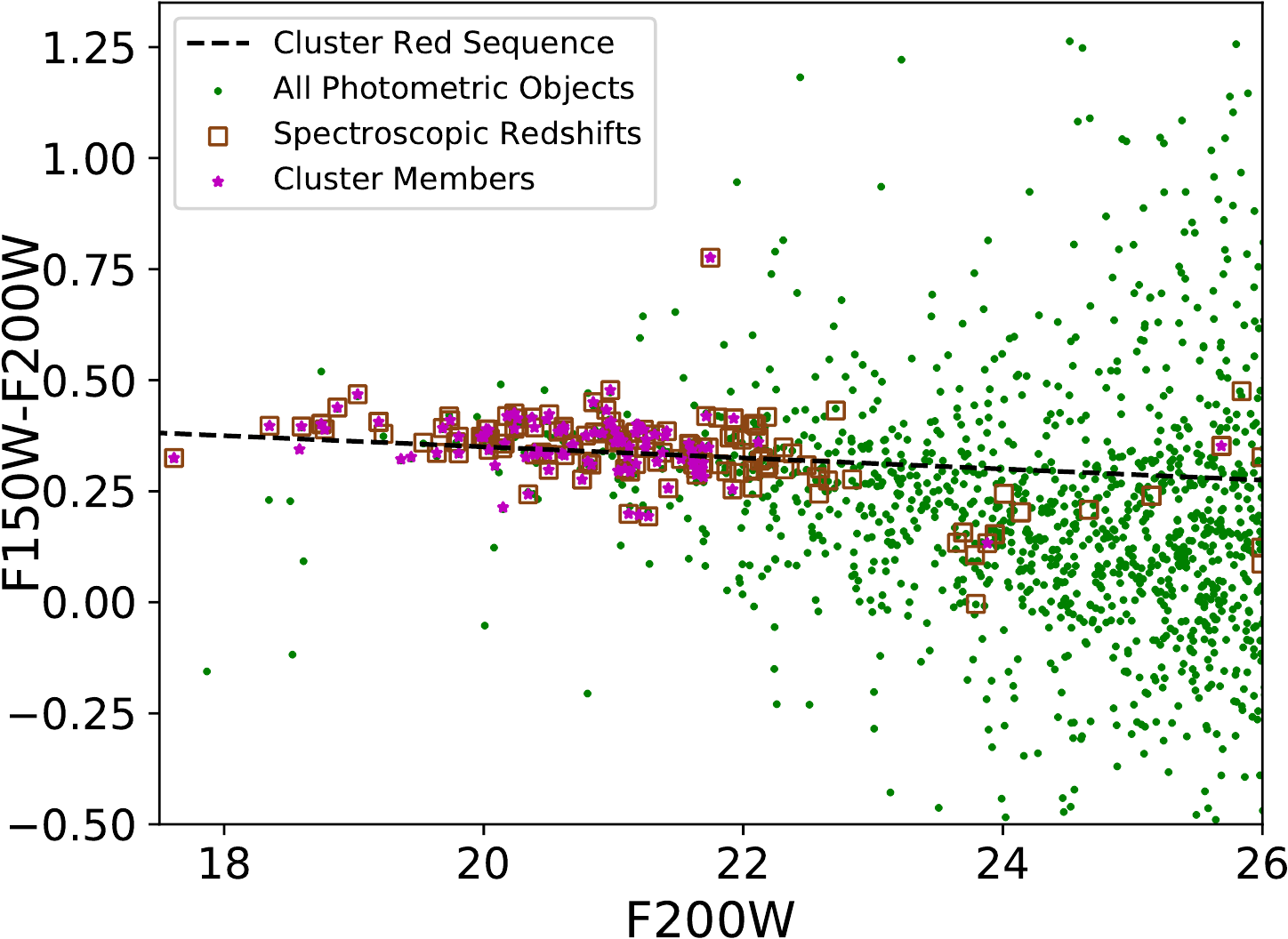}
\caption{ 
Color--magnitude diagram generated from the JWST/NIRCam photometry. Cluster members are indicated by red dots, and other galaxies are indicated by green dots.  Galaxies with spectroscopic redshifts are surrounded by squares. The black dashed line shows the cluster red sequence. Magnitudes are SExtractor AUTO, and colors are based on PSF-matched isophotal photometry.}
\label{fig_CMD}
\end{figure}

\subsection{Full object catalog} \label{sec:phot_catalog}
Two independent software approaches were applied to estimate photometric redshifts: EAZY \citep{Brammer2008,Brammer2021} and LePhare \citep{Arnouts2011}. In both cases, SED templates were optimized for identification of high-redshift galaxies using JWST/NIRCam imaging \citep{Larson2022}.
Comparison of 161 galaxies that have both spectroscopic and photometric redshifts shows good agreement for both LePhare and for EAZY\null. In 98\% of cases, the photometric redshifts are  within 15\% of the spectroscopic redshifts.  There are nevertheless a handful of outliers (Figure~\ref{fig_photoz}).  Most of the outliers are at redshifts at which strong nebular emission lines are  between two filters. For example, at $z=1.1$ the observed wavelength of H$\alpha$ is  between the F115W and F150W filters, and at $z=2.6$ H$\alpha$ is  between the F200W and F277W filters.  In these cases, this strong nebular emission line that would increase the flux within a filter gets missed and therefore cannot contribute to the redshift measurements. 
Although the LePhare redshifts have two outliers that  EAZY  doesn't,  the LePhare redshifts show slightly less scatter at $z=4.3$ where they are most important for this study. For this reason we  adopted LePhare for the remainder of this study, but given  the similar performance between the two photometric redshift approaches and the many spectroscopic redshifts, the choice should make little difference.

After the successful checks, we extended the photometric redshift estimates to the full multi-band object catalog.   A photometric redshift is considered secure if the object is: (1) in the field-of-view for all filters, (2)  detected in a minimum of eight filters, and (3) spatially resolved from its neighbors. The resulting distribution of photometric redshifts peaks at the cluster redshift and displays also minor peaks $z\approx1.6$, $z\approx2.4$, $z=3.75$, and  $z=4.32$ (Figure~\ref{fig_photoz}). The $z\approx1.6$ peak is caused by a problem case in which [\ion{O}{3}] falls in a gap between the F115W and F150W filters and at the same time H$\alpha$ is situated between the F150W and F200W filters, resulting in redshift degeneracies.  Likewise, at $z\approx2.4$ [\ion{O}{3}] falls between F150W and F200W. The $z=4.32$ peak corresponds to the redshift of the known galaxy overdensity mentioned in \S\ref{sec:intro}.

\begin{figure}[h]
\centering\includegraphics[scale=0.40]{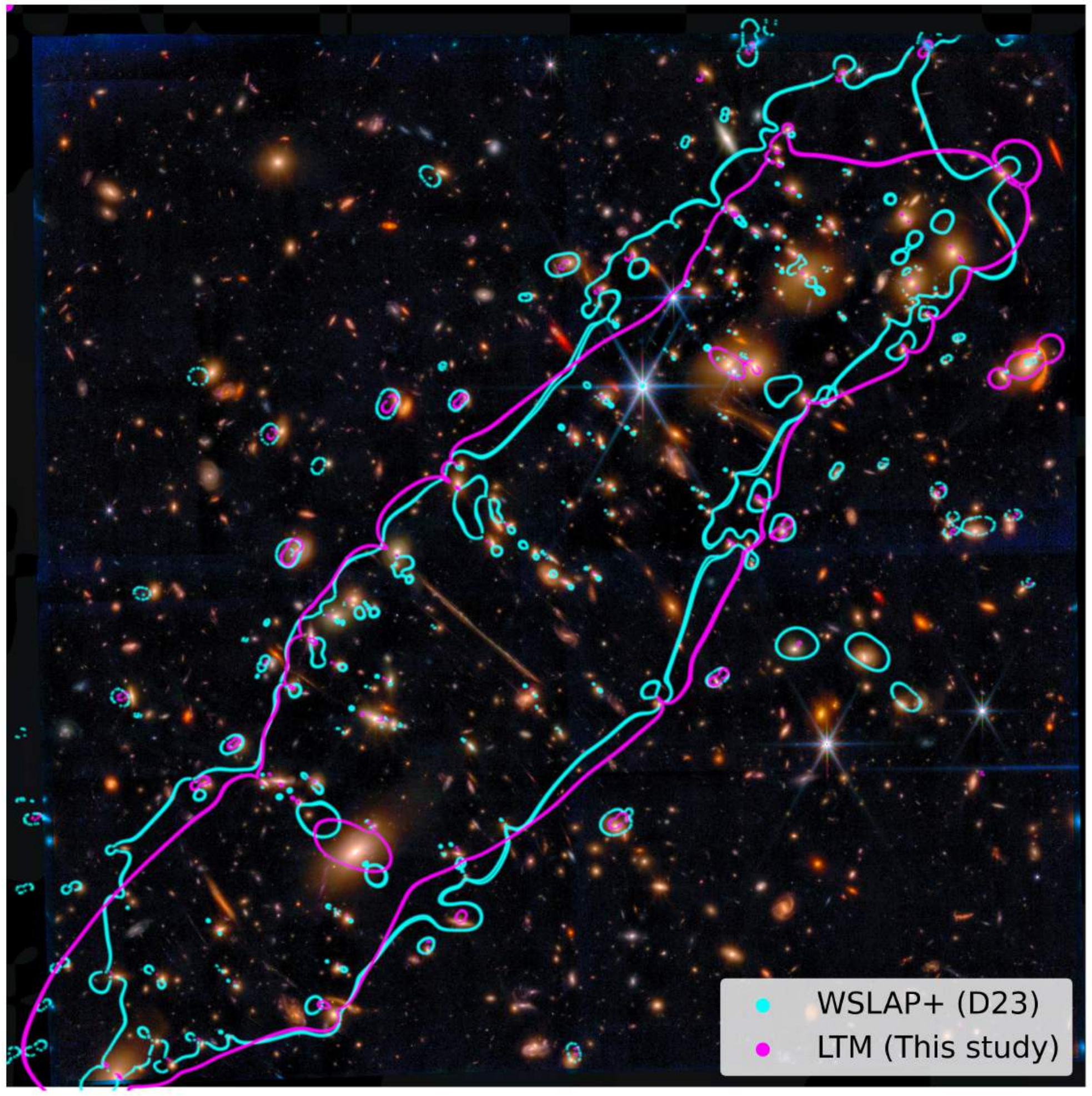}
\caption{Color image using all NIRCam filters that depicts the critical curves for our LTM model at $z=4.3$ (magenta) and for the free-form WSLAP+ model at $z=2.5$ \citep[cyan;][]{Diego2023}. The critical curve is depicted at a $z = 4.33$ in order to match the redshift of the galaxy overdensity discussed in \S7. Orientation and color rendering are the same as Figure~\ref{fig_map}. 
}
\label{fig_lens}
\end{figure}

The redshift peak at $z=3.75$ is poorly understood. The spectroscopy found only one lensed source at a redshift near this peak ($z=3.77$), yet the probability distributions of the photometric redshifts of the other sources are largely consistent with a single peak and do not allow for other redshift solutions. A concern is that an instrumental gap in the MUSE spectrograph prevents the detection of features in the wavelength range 5805--5965~\AA, corresponding to Ly$\alpha$ at $z=3.75$. Nonetheless broader features such as the Ly-series continuum break should  be detectable if there really is a galaxy group at this redshift. This problem is compounded by the fact that at this same redshift, the H$\alpha$+[\ion{N}{2}] complex falls between the F200W and F277W bands and so is not recorded, while H$\beta$+[\ion{O}{3}] falls within F200W, resulting in redshift degeneracies \citep{McKinney2023}. These special circumstances explain why some sources at $z=3.75$ might be missed but do not explain a peak at this redshift. In sum, the peak in lensed source counts at $z=3.75$ may indicate the presence of a bona fide galaxy overdensity, but its existence remains uncertain pending spectroscopic confirmation.

\section{Strong Lensing Model} \label{sec:LTM}

\subsection{Light-traces-mass Approach}
We constructed a {light-traces-mass} ({LTM}) strong-lensing model, which takes advantage of strong-lensing evidence and especially of image systems \citep{Zitrin2009, Zitrin2015}. This model approach requires few free parameters and therefore minimizes overfitting of the lensing constraints and resulting unphysical solutions. The model began with the central cluster galaxies, each with an assumed 2D power-law surface-mass density profile of power-law index $q$ and with mass normalized by the measured luminosity of each galaxy, to create a cluster galaxy mass map in 2D\null. An approximate ``cluster" mass component was generated by summing up the galaxy mass distributions and smoothing with a Gaussian kernel of width $S$. These galaxy and cluster components were then added with a relative weight $k_{\rm gal}$, reflecting the ratio of luminous to dark matter, which together were scaled to a desired redshift by a factor $K$. A third component is a global external shear of strength $\gamma_{\rm ex}$ and position angle $\phi_{\rm ex}$. The external shear allows for elongated critical curves but contributes only to the deflection field and not to the total mass density. Thus the model's six free parameters are $q,S,K,k_{\rm gal},\gamma_{\rm ex}$, and $\phi_{\rm ex}$.  When fitting images that lack a secure redshift measurement, the source distance is an additional free parameter to be optimized in the fit. The best fit that satisfies the free parameters was obtained by a Markov Chain Monte Carlo (MCMC) approach with uncertainties computed by bootstrapping the MCMC steps. We refer to \citet{Pascale2022a} for details on the implementation of this code.


The cluster members inputted to the model were selected by their spectroscopic redshifts and also by making a strict color cut to isolate the red sequence (Figure~\ref{fig_CMD}). To correlate the mass with the luminosity, the cluster members were all nominally assigned the same global weight factor of 1.0, which was subsequently modified for two special regions. First, a galaxy overdensity in the direction of the SE component in the foreground at $z\approx0.63$ introduces a non-negligible deflection to the lens model.  Following previous work \citep{Caminha2022,Diego2023}, we adopted the cluster redshift for these galaxies but assigned a weighting factor  less than unity to account for the different lensing geometries. Second, a handful of galaxy members which are ultra-bright and/or situated near in projection to the critical curve have a higher lensing impact than their individual brightnesses would imply. As such, the weighting factors and configurations (core radius, position angle, and ellipticity) were left free to be fit by the model. These galaxies include the brightest cluster galaxy (BCG), which is situated in the SE component, and other influential members such as the galaxy closest in projection to El Anzuelo (system 24 in Figure~\ref{fig_map}).

\begin{figure}[tb]
\centering\includegraphics[scale=0.29]{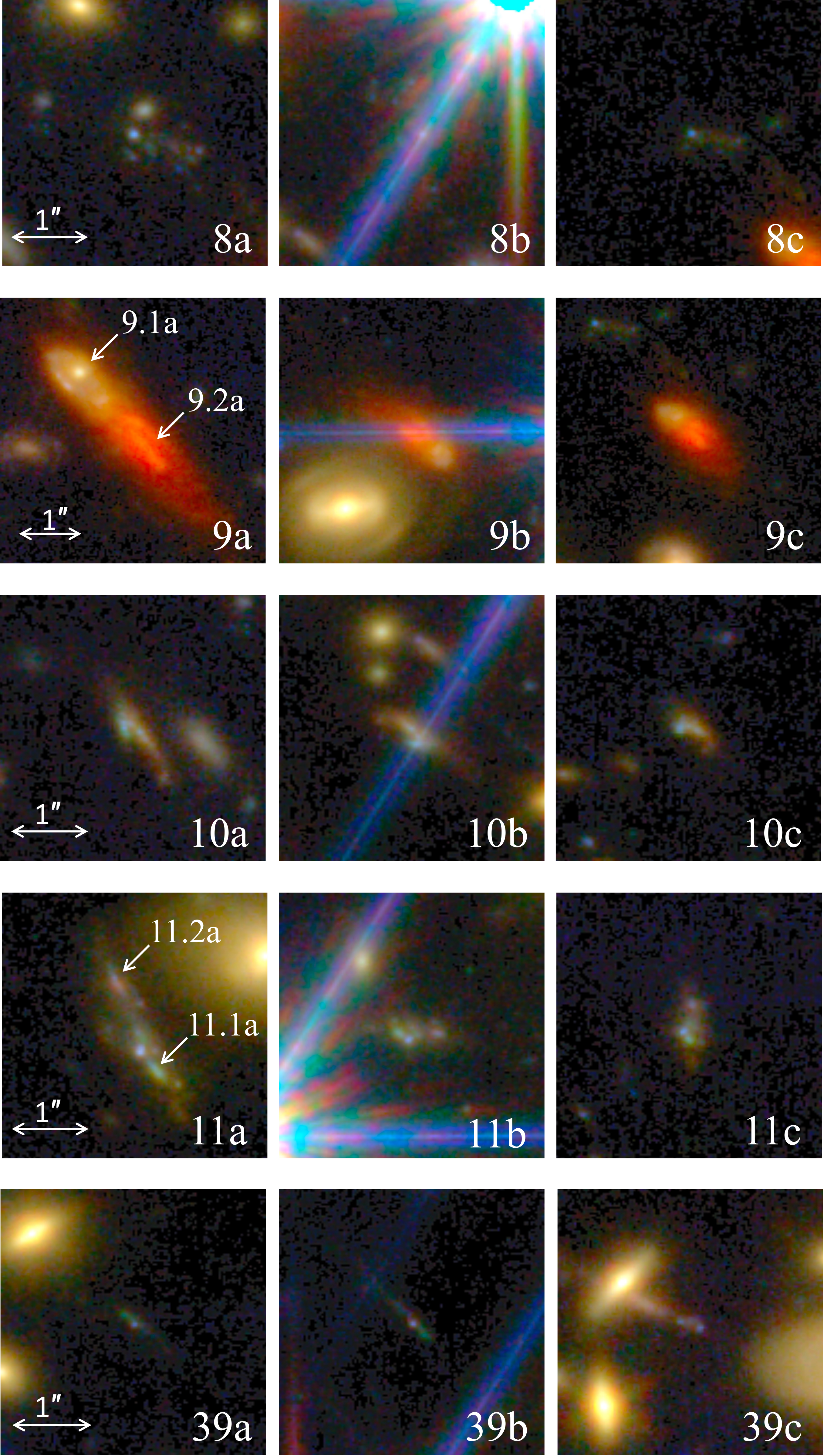}
\caption{NIRCam color images of the four known and one probable $z=4.32$ systems showing resolved morphologies.  Scale bars are shown in each image, and the orientation is North up, East left.  Systems 8, 9, 10, and 11 were first identified by \citet{Caputi2021}, in whose study the HST imaging was not able to detect the red component in system 9. In NIRCam the galaxy images separate out into multiple components, suggesting ongoing galaxy interactions/mergers, and also flip in image parity between the counterimages, as expected on each crossing of the critical curve. System~39 was discovered in this study by its model-predicted redshift, photometric redshift, and morphology. 
}
\label{fig_z4p3}
\end{figure}

\begin{figure*}
\centering\includegraphics[scale=0.42]{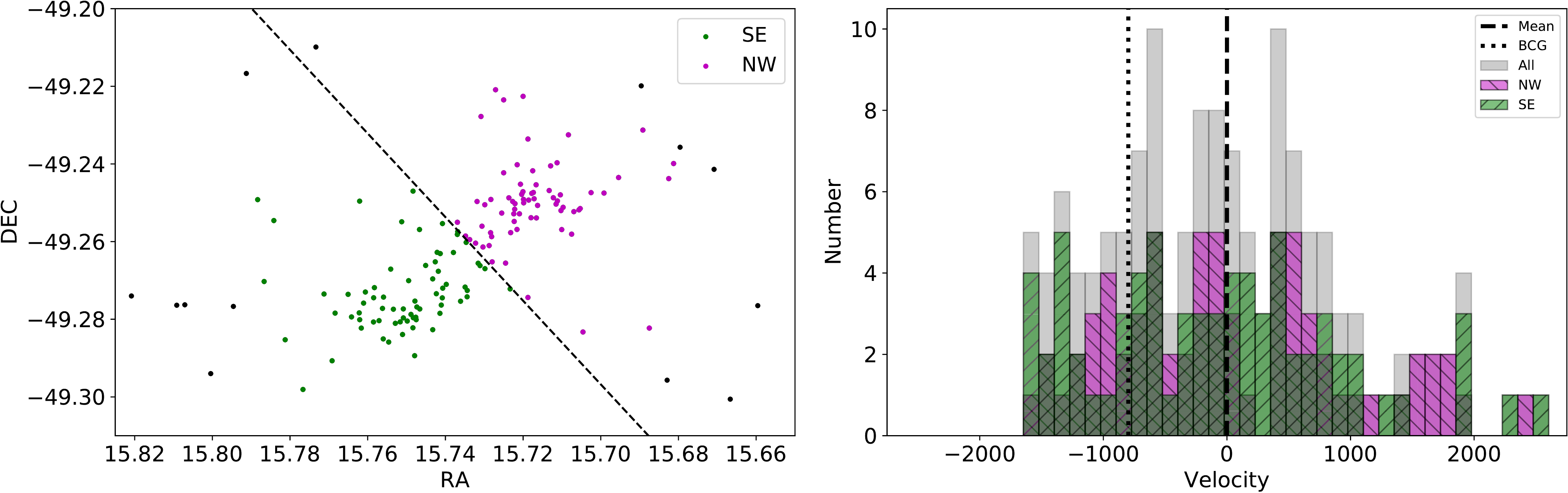}
\caption{Spectroscopy in the El Gordo cluster field. {\it Left}: The cluster members with spectroscopic redshifts are depicted by the colored dots.  Cluster members are assigned a different color with respect to a cut that bisects the cluster (black dashed line). {\it Right}: the redshift histogram is depicted for all cluster members (gray), and separately for the NW (purple) and SE (green) sides. The dashed vertical line corresponds to the mean cluster redshift of $z=0.873$, and the dotted line gives the relative velocity of the BCG. The components of this ongoing major cluster merger do not obviously separate out in velocity space, motivating a more complex statistical analysis of the galaxy velocities to uncover any substructures.}
\label{fig_histo}
\end{figure*}

The lensed-galaxy constraints were  introduced in a multi-step process, starting with the 23 image systems with spectroscopic redshifts \citep{Caminha2022}. We also added El Anzuelo, which is a partial Einstein ring and ALMA source for which CO(3--2) line emission is detected at $z=2.291$ \citep{Diego2023, Kamieneski2023}. 
%
The initial lens model was based on these 24 spectroscopically-confirmed systems with their redshifts fixed. Parameters of the remaining systems were free to be fit by the model. Additional secure image systems were introduced gradually, each time making sure that the fit improved. To qualify as secure, we required that all images in a given image system must have similar colors and spectroscopic redshifts (where available), a consistent lens-predicted location and model-predicted redshifts, and correlated morphologies (for uncontaminated images). Based on our criteria, we were unable to accept six image systems from the literature.  The Appendix gives details. We augmented the list of image constraints in stages:  (1) new image family members of known image systems, and (2) new image systems, which were all subsequently vetted by our lens model. In total we identified five new counterimages of known image systems and two new image systems (61 and 62). The image systems from \citet{Caputi2021} are reported in Table~\ref{tab_1}, and the full set of lensing constraints is reported in Table~\ref{tab_2}. The image system identifications for the first 23 systems follow the designations of \citet{Caminha2022}, and the remainder follow the designations of \citet{Diego2023}. 

Our best fit model, which is the one for which $\chi^2$ is minimized, includes a total of 56 image systems and is presented in Figure~\ref{fig_lens}.  This model reproduces the angular positions of input lensed images to an rms difference of $\sim$1\farcs8.  
It is reassuring that the critical curve has a similar orientation, shape, and Einstein radius in both the LTM and WSLAP+ models.
Although the new model incorporated most of the same image systems as \citet{Diego2023}, our LTM model was constructed independently and by a different method. 

The LTM lens model gives a total mass $(7.0 \pm 0.3) \times 10^{14}$~\Msol\ within a radius of 500~kpc of the cluster's luminosity-weighted mean center (LWMC; see \S5). The uncertainty is dominated by systematics \citep{Johnson2016, Meneghetti2017, Strait2018, Pascale2022a} because strong-lens modeling is an inherently underconstrained problem. Systematic uncertainties estimated above come from two sources: differences in galaxy mass-to-light ratios (assumed by the LTM approach to be identical for all galaxies) and uncertainty in fitting model parameters to the data.  Each uncertainty was estimated by an MCMC approach using 100 iterations on a simplified model using only the 24 image systems having spectroscopic redshifts. The 1$\sigma$ uncertainty from 0.3~dex variation in the mass-to-light conversion factor is $1\times 10^{13}$\,\Msol\ or 2\% of the inferred mass, while the model construction itself gives an uncertainty of $8\times 10^{12}$\,\Msol\ or  1\%. These uncertainties are small but consistent with those of \citet{Johnson2016}, who concluded that lens models based on at least one spectroscopic image system yield uncertainty $<$4\% in mass within 1~Mpc. To be conservative, we adopted 4\% as the overall uncertainty.

The mass given by our LTM model is slightly smaller than that in 
\citet[][]{Diego2023}, who estimated a mass of 8.0--$8.6\times10^{14}$\,\Msol\ in the same 500~kpc radius we used.
Our values may be consistent with an older LTM model \citep{Zitrin2013}, based on nine image systems, none  with spectroscopic redshifts, for which the mass was reported to be $6\times10^{14}$~\Msol\ in a 370~kpc radius and $1.7\times10^{15}$~\Msol\ in 700~kpc.
At the same time, the weak-lensing mass within a radius of 0.8 Mpc for the NW and 0.7 Mpc for the SE components that roughly corresponds to our radius of 500 kpc from the LWMC, is   $M_{500} = (1.8\pm 0.34)\times$\,10$^{15}$\,\Msol\  \citep[][their Table 2]{Jee2014}.  Their value is comparable to the high-end of the strong-lensing values. 

\subsection{Spectroscopic Image Systems at $z=4.32$} \label{sec:odense1}
Four galaxies at $z=4.32$ are triply-imaged and previously-known \citep{Caputi2021}: image systems 8, 9, 10, and 11 (Figure~\ref{fig_z4p3}). We recount below the photometric, lensing, and morphological information that is new to this study, and report redshift values which are sourced from \citet{Caputi2021}. To start, Arc 8a has a spectroscopic redshift of 4.3175. With only NIRCam data, Arcs 8a and 8c gave  $z_{\rm ph}\approx0.7$ \citep{Diego2023},  likely  a result of misidentification of the Lyman break as the 4000~\AA\ and Balmer breaks. The inclusion of the bluer HST bands helps to break this degeneracy, yielding a better photometric redshift estimate for these two galaxy images. Arcs 8a and 8c consist of  multiple components. The brightest clump near Arc 8a (yellow knot above the blue knot in Figure~\ref{fig_z4p3}) is of concern because it does not appear in Arc 8c and has the colors of a cluster member.  Therefore, the SED  was measured by integrating over the  bluer and more compact arc clumps,  excluding this apparent cluster member. Our lens model gives magnification factors $\mu$ of 4.2 and 3.1 for Arcs 8a and 8c, respectively.  Arc 8b is severely contaminated by a bright stellar diffraction spike, and no useful photometry is possible.

Arc 9.1a has  $z_{\rm sp}=4.3196$. Each image of this triply-imaged system has two main components: a bluer one  detected  in the HST imaging (Arcs 9.1a,b,c) and a redder one detected  in archival ALMA imaging (Arcs 9.2a,b,c).  Both components appear in the NIRCam imaging. Our photometric redshift easily recovers the spectroscopic redshift for Arc 9.1a, which is detected in every NIRCam filter. Arc 9.2a, on the other hand, has lower signal to noise and is not detected in the SW NIRCam filters; it yields no photometric redshift. This dual-colored lensed image flips in parity between the counterimages, as expected on each crossing of the critical curve. The lens model gives $\mu\approx 3$--8, depending on the component (Table~\ref{tab_2}).

Image system 10 has spectroscopic redshifts measured individually for all three images:  $z_{\rm sp}=4.3275$ for Arc 10a, $z_{\rm sp}=4.3269$ for Arc 10b, and $z_{\rm sp}=4.3289$ for Arc 10c.  Similar to system 8, the NIRCam-only $z_{\rm ph}\approx0.7$ measured for Arcs 10a and 10c in \citet{Diego2023} is incorrect, and adding HST data produces photometric redshifts consistent with the true ones.  The morphology resembles an upside-down blue ``check" and a red mark.  This ``checkmark" galaxy image  flips in parity twice across the three images, as expected for each crossing of the critical curve.  We estimate magnification factors of 4.0 and 3.0 for Arcs 10a and 10c, respectively.  Arc 10b suffers from contamination by a bright stellar diffraction spike.

Image system 11 has spectroscopic redshifts measured individually for all three images:  $z_{\rm sp}=4.3278$ for Arc 11.1a, $z_{\rm sp}=4.3273$ for Arc 11.1b, and $z_{\rm sp}=4.3273$ for Arc 11.1c. Again the NIRCam-only photometric redshifts are unreliable with  $z_{\rm ph}=0.69$ for Arc 11.1a and $z_{\rm ph}=4.11$ for Arc 11.1c. Photometric redshifts using our HST+NIRCam catalog  are consistent with $z=4.32$ for both images. The morphology consists of multiple blue and red clumps which again appear to be indicating galaxy--galaxy associations or interactions.   Arc 11.1b has severe contamination from a stellar diffraction spike, and no photometric redshift could be obtained. Magnification factors range from 2.9 to 4.5.

\begin{figure}
\includegraphics[viewport= -5 0 350 755,scale=0.48]{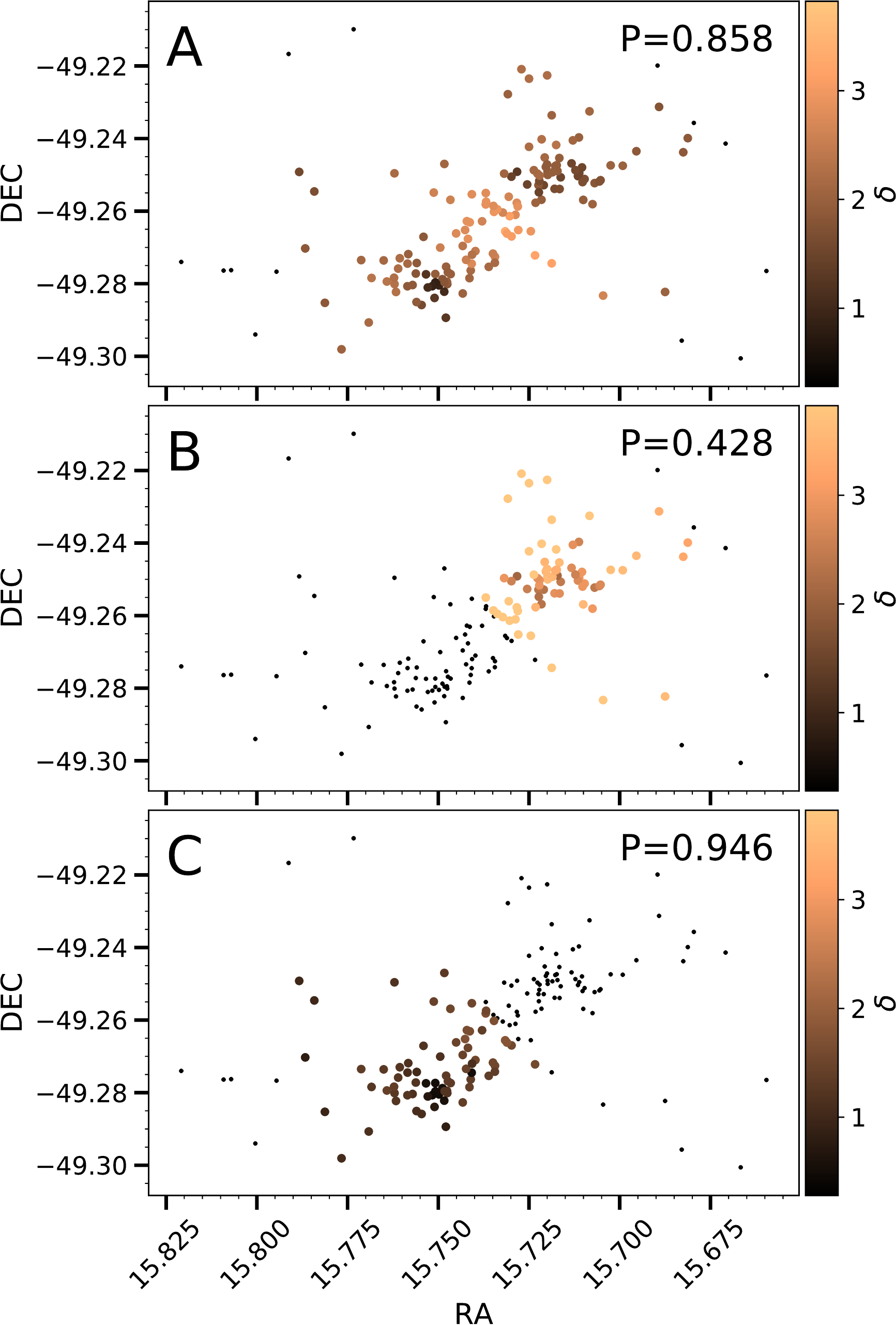}
\caption{Search for substructure in El Gordo.  Each dot, located at the position of a cluster member, represents the $\delta$ value computed using the 12 nearest neighbors. The dots are color-coded relative to the $\delta$ values as indicated in the color bars.   In  this rendition, darker colors indicate the presence of more tightly-bound regions.
Values of $P$ are recorded in each panel, for which higher values correspond to a more dynamically-bound structure. {\it Panel A:} $\delta$ for all galaxies in the cluster.
 {\it Panels B, C:} $\delta$ calculated separately for the two cluster components.}
\label{fig_delta}
\end{figure}

\section{Spectroscopic Analysis} \label{sec:spec}
Spectroscopic redshifts for El Gordo  ($z=0.873$) are drawn from two main sources. First, \citet{Sifon2013} obtained redshifts for 89 galaxies which  extend  out to 2.5~Mpc (physical), well outside of the NIRCam field-of-view. Second, \citet{Caminha2022} secured redshifts for 402 objects in the central 1\arcmin$\times$3\arcmin\ region of the cluster. The redshifts range from 0 up to 5.9521 with fairly uniform azimuthal coverage owing to the IFU spectroscopic setup.  In all they found 150 galaxies near the cluster redshift ($0.85<z<0.90$).

Cluster membership is met for galaxies which have velocities within $\pm$4000~km~s$^{-1}$ of the mean value of $z=0.873$.  This corresponds to the range $0.862 < z < 0.890$. We further selected the subset of galaxies within the projected virial radius of $1.7$~Mpc from the cluster center. By these criteria, a total of 140 galaxies make it into our  cluster-member catalog. Figure~\ref{fig_histo} (left) shows the galaxy positions. 

The number of cluster galaxies with radial velocities suffices to assess cluster disturbances \citep[\eg][Appendix~A]{Windhorst2018} and also  to estimate the mass. In the simple approximation that a single dark matter halo underlies the cluster, the virial theorem applied via the Gapper method \citep{Wainer1976,Beers1990} yields $M_{1}= (4.6\pm0.54)\times10^{14}$~\Msol\ within the virial radius (1.7~Mpc). (The uncertainty was computed by jackknife sampling following the prescription of \citealt{Beers1990}.) However, El Gordo is a double cluster, and that has to be taken into account.
For a double cluster with mass ratio unity and separation between components $\epsilon$ times the virial radius, the above $M_1$ underestimates the mass by a factor of $(1-\frac{7}{16}\epsilon ^2$). For El Gordo, $\epsilon \approx 0.40$, resulting in a correction factor of 0.93, and total mass $M_2 = (5.1\pm0.60)\times10^{14}$~\Msol. 
The stated uncertainty is based solely on the uncertainty in the velocity dispersion.
$M_2$ is smaller than the mass reported by \citet{Menanteau2012},  $1.86^{+0.54}_{-0.49} \times 10^{15}$\,\Msol.  Those authors also reported a mass ratio closer to 2:1 between the NW and SE components while ours is closer to unity (\S6). \citeauthor{Menanteau2012} estimated the mass in a different way, applying  a scaling relation between $M_{200}$ and dark matter concentration to the velocity dispersion, which incurs additional uncertainty. 


$M_2$ is about a factor of 4 smaller than the virial mass estimated by a strong-lensing approach to similar radius, 2.09--2.24$\times$10$^{15}$ \citep{Diego2023}. A recent weak-lensing estimate also to similar radius of 2.13$^{+0.25}_{-0.23}$$\times$10$^{15}$~\Msol\ \citep{Kim2021} is similar to the strong-lensing mass and larger than $M_2$.
This discrepancy with the strong- and weak-lensing results can be explained in part by orientation effects.  Absence of a double peak in the redshift histogram suggests that the cluster is oriented near the plane of the sky with a significant transverse velocity component.
If this is the situation, the true virial mass will be larger than $M_2$ above. Other factors that can bias the mass include the existence of other cluster halos/subhalos and the validity of the assumption of hydrostatic equilibrium for a merging cluster.



El Gordo is observed during a cluster merger involving at least two components, thereby explaining its elongated structure. 
Even though the double-peaked galaxy distribution is well known, there is no obvious bimodality in the radial velocities (Figure~\ref{fig_histo} right), similar to the results of \citet{Menanteau2012}.  
If the collision axis is the line drawn through the centers of the SE and NW components, then bisecting that  line (Figure~\ref{fig_histo} left) and computing the velocity histograms on each side still does not uncover two distinct components.  However, the mean radial velocity difference between the 71 galaxies southeast of the dividing line and the 69 galaxies on the northwest side is $\sim$300~km~s$^{-1}$ in the cluster rest frame. If the collision axis is in the plane of the sky, the transverse velocity difference should be much larger than that.

Building on the work of \citet{Dressler1988}, \citet{Biviano2002} devised a statistic to search for galaxy substructure.  In their approach, the parameter $\delta$ compares the local radial velocity properties of the Student-$t$ and $\chi^2$ distributions with their global values. For a cluster with $N$ redshifts, the group size is $\sqrt N$, equating to 12 nearest neighbors for this study.  Figure~\ref{fig_delta} depicts the distribution of $\delta$ values, with each point representing the $\delta$ value computed from its dozen nearest neighbors.  In turn, the individual 140 $\delta$ values are summed up to yield $\Delta_{\rm obs}$. To calibrate this statistic, a final step  compares the sum of the $\delta$ values of one thousand randomly generated instances in which the redshifts of each galaxy remain the same  but their positions are azimuthally scrambled, $\Delta_{\rm sim}$. For each instance in which the data have a smaller and more bound $\Delta_{\rm sim}$ value, the numerator of the probability $P$ relative to 1000 increments by one. From this exercise, we calculate the value $P$ for which the randomly generated $\Delta$ will be larger than the  observed $\Delta_{\rm obs}$. 

To get an intuition for how to interpret these values, in a study of 59 clusters \citet{Biviano2002} found $P \leq 0.05$ to indicate significant substructure. On the other hand, a  value of $P$ closer to 1.0 indicates a more dynamically intact structure that lacks significant substructures. For El Gordo, $P = 0.858$ (Panel A of Figure~\ref{fig_delta}).  The probability is thus fairly high that the cluster as a whole is bound.  To gain more physical insights, we isolated the cluster into the NW and SE halves and recomputed $P$. We obtain $P= 0.428$ and 0.946, for the NW and SE components, respectively (Panels B and C of Figure~\ref{fig_delta}. Notably, $P$ approaches 1.0 only for the SE component, consistent with that component having a higher mass density in the vicinity of the BCG\null.


\section{El Gordo Physical Properties} \label{sec:props}
 
%
%
%

\begin{figure}[h]
\centering\includegraphics[scale=0.24]
{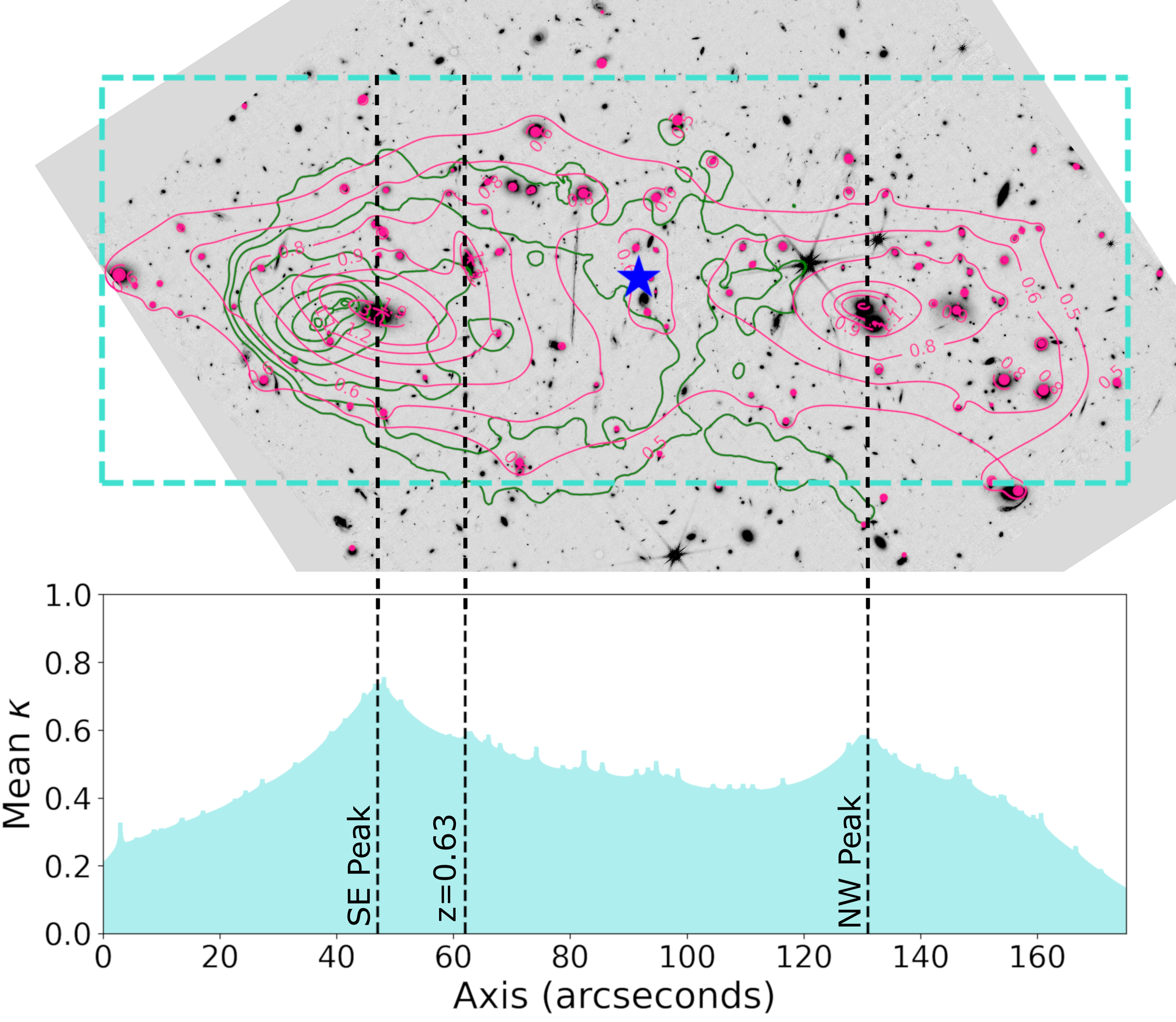}
\caption{Surface mass contours scaled to the critical value ($\kappa$) from our lens model  (red continuous contours), relative to the X-ray map (green continuous contours). The mass peak is offset from the positional centroid in the X-ray map by a physical separation of between 5 and 9 kpc, similar in extent to the nonparametric lens model using NIRCam data by \citet{Diego2023}. The X-ray peak is correlated with the position of a bright filament thought to be associated with a dense cooling flow onto the BCG.  The lower panel depicts the one dimensional scan of $\kappa$ summed up in a column orthogonal to the collision axis, obtained by summing up $\kappa$ over an angular range of $\pm$35$^{\prime \prime}$, equating to $\pm$275 kpc along the at $z=0.87$. The SE and NW component mass peaks are detected, as is an intervening mass peak at $z=0.63$ that is taken into account in our lens model. The integrated mass does not have a low minimum, suggesting a significant amount of mass near the position of the LWMC (blue star-shaped symbol).}
\label{fig_kappa}
\end{figure}

The mass distribution gives insights  into  physical properties of this cluster. In a one-dimensional trace of the surface mass density (Figure~\ref{fig_kappa}), both mass peaks are prominent, and there is a smaller peak at $z=0.63$ situated between the two major components.  A depression appears between the peaks in the 1-d trace (lower panel of Figure~\ref{fig_kappa}) but not a sharp mass cut-off.  The masses on the two sides of the cluster (divided at the LWMC position: blue star in Figure~\ref{fig_kappa}) are ($3.6 \pm 0.15) \times 10^{14}$\,\Msol\ for the SE component and
($3.4 \pm 0.15) \times 10^{14}$\,\Msol for the NW\null. 
This makes the SE component slightly more massive than the NW one.
Our statistical analysis (Panel~c in Figure~\ref{fig_delta}) makes the SE component more tightly-bound, which may yield a higher galaxy number density there.

Other strong- and weak-lensing models also slightly favor the SE component \citep{Zitrin2013,Cerny2018,Diego2020,Kim2021,Caminha2022,Diego2023}, even though earlier studies had favored a NW-dominant mass with mass ratio $\sim$0.6:1 \citep{Menanteau2012,Jee2014}.  Even those studies, though, were consistent with 1:1 within the uncertainties.  The angular separation between the SE and NW components is 83\arcsec\ equating to 650~kpc at the redshift of the cluster. If the transverse velocity is similar to the transverse velocity dispersion (\S\ref{sec:spec}), the crossing time is $\sim$1.3~Gyr. In the more likely event the transverse velocity is higher, the time would be correspondingly shorter.

NIRCam detects not only gravitationally-lensed arcs in the central region of the cluster but also filaments of a cooling flow, which are somewhat arc-like in appearance yet lack the morphology and geometry of lensed background-galaxy images \citep[Figure~\ref{fig_map}; see also][]{Diego2020,Caminha2022}. The  X-ray source (Figure~\ref{fig_kappa}) has a cometary shape and is mostly confined to the SE component despite the relatively equal mass ratio between the two components discussed above and their presumed similar  dynamical history. On closer evaluation, the X-ray peak is a bit elongated, offset from the BCG by 5--9~kpc, and centered on the cooling flow of the BCG\null, suggesting a physical association. The interpretation that the cooling flow is mainly responsible for the X-ray emission is also motivated by the high particle number-density in cooling flows, such that the bremsstrahlung radiation within the cooling flow is stronger than the same radiation mechanism operating within the more diffuse and much more widespread intracluster medium. 

Because the X-rays might not be a reliable indicator of the global cluster properties, we can turn to the spectroscopy. For example, then velocity offset of the BCG from the systemic redshift of the cluster is a known diagnostic of cluster virialization \citep{Rumbaugh2018}. 
For El Gordo, the BCG has a measured velocity offset of $\sim$800 km~s$^{-1}$, from which we infer that the cluster is not well approximated as a virialized system (Figure~\ref{fig_histo}).  Moreover, the centroid position of the projected luminosity of the spectroscopically-confirmed cluster members is another known diagnostic of cluster virialization \citep{Rumbaugh2018}. For our case, the ``LWMC" was computed from the F200W image.  Its rest wavelength of 1.07~\micron\ is a proxy that samples the SED longward of the 4000~\AA\ and Balmer breaks at the cluster redshift. The resulting LWMC (Figure~\ref{fig_kappa}) is situated near the midpoint of the two cluster components rather than near the BCG\null.  This corroborates that there has been a major disturbance of the cluster.

At longer wavelengths, data obtained at 610~MHz and 2.1~GHz using the Giant Metrewave Radio Telescope and the Australia Telescope Compact Array uncover two filamentary structures of radio emission which are aligned along the collision axis, positioned on diametrically-opposite sides, and separated by the viral radius \citep{Linder2014}.  These ``radio relics" are thought to be synchrotron emitting regions induced by shock waves propagating through the intracluster medium during mergers between two galaxy clusters. According to $N$-body simulations of the cluster-cluster merger, which are constrained by the X-ray, SZE, HST lensing, and dynamical data, the most likely scenario is that El Gordo is a simple binary cluster that made its first pericentric passage $\sim$480 Myr ago \citep{Molnar2015}. It is not yet established whether the collision was head-on or off-axis \citep{Zhang2018}. A dark-matter-only Monte Carlo modeling code applied to these same constraints finds that the SE and NW components may be inbound \citep{Ng2015}, which would naturally explain the radio relics trailing the two subhalos, each of which has a synchrotron lifetime $\apll$10$^{7.5}$~years, {\ie}, considerably shorter than the crossing time measured above.  As a caveat,  \citet{Kim2021} used Monte Carlo simulations and the radio relic constraints to show that neither the outgoing nor return phase is fully supported by the available information.

In sum, merger simulations, our study, and other studies  have obtained several new constraints on the merger properties, 
such as: (1) the LWMC is situated near the midpoint of the two components; (2) the line-of-sight peculiar velocity difference between the merger components is $\sim$300~km~s$^{-1}$;  (3) the velocity difference between the BCG and the NW component is $\sim$400~km~s$^{-1}$;  (4) the $\kappa$-map has a center near the BCG and is offset from the X-ray peak, consistent with the cooling flow being the primary source of the X-ray emission; (5) the mass ratio is near unity 
(6) the crossing time is large relative to the lifetime of the radio relics.  New simulations which take these constraints into account can help to place this massive and high-redshift cluster into the larger context of mass assembly.

\section{Background Galaxy Overdense Region}  \label{sec:odense}

\subsection{Background}
\citet{Zitrin2013} identified a triply-imaged system with a model-predicted redshift of 4.16. This system corresponds to Arcs 10a, b, c. \citet{Caputi2021} measured a spectroscopic redshift for this arc of $4.33$ and measured similar redshifts for three other systems (here 8, 9, and 11---Table~\ref{tab_1}). By fitting SEDs to the rest-frame UV data, \citeauthor{Caputi2021} found all members to be star forming and to have low  dust extinction. Two of the galaxies have relatively high specific star formation rates, from which they inferred that these galaxies may be undergoing galaxy--galaxy interactions. 

\begin{deluxetable}{ccccc}
\tablecaption{Galaxy Overdensity Members ($z=4.32$)}
\tablecolumns{8}
\tablehead{
 \colhead{ID} &  \colhead{$z$} & 
  \colhead{$z_{\rm ph}$} & \colhead{$\mu$} & \colhead{$m_{\rm F200W,int}$}  
}
\startdata
8a  &4.3175$^a$&$4.12^{+0.08}_{-0.09}$& 4.2 &26.84  \\
8c  &4.3175$^c$&$4.01^{+0.26}_{-0.15}$& 3.1& 27.16\\
\hline
9.1a&4.3196$^a$&$4.47^{+0.02}_{-0.02}$& 5.6 &24.07 \\
9.1c&4.3196$^c$&$4.29^{+0.09}_{-0.08}$& 3.0 &25.40 \\
9.2a&4.3196$^a$&  & 7.5 &\no\\
9.2c&4.3196$^c$&  & 2.9&\no  \\
\hline
10a&4.3275$^a$ & $4.42^{+0.07}_{-0.05}$ & 4.0&25.57 \\
10b&4.3275$^a$ &  & 3.4&$>$27.93\\
10c&4.3275$^a$ &$4.29^{+0.06}_{-0.08}$&3.0& 25.90 \\
\hline
11.1a&4.3278$^a$& $4.27^{+0.07}_{-0.06}$& 4.5&25.21 \\
11.2a&4.3278$^b$&$4.29^{+0.06}_{-0.05}$    & 4.2&25.82 \\
11.1b&4.3273$^a$&  & 3.0& \no \\
11.2b&4.3273$^b$& &2.9 &\no \\
11.1c&4.3273$^a$&$4.34^{+0.09}_{-0.07}$ &  3.2 &25.78 \\
11.2c&4.3273$^b$&     & 3.0 &\no \\
\hline
39a &4.14$^d$&$4.29^{+0.16}_{-0.16}$ & 3.8  &27.28 \\
39b &4.14$^d$&                       & 7.2  &26.23 \\
39c &4.14$^d$&                       & 3.3  & \no \\
\hline
101 &$4.317^e$ & &2.1 &{$>$28.77} \\
\hline
102 &4.32$^f$&$4.12^{+0.24}_{-0.13}$& 1.5 & 26.78 \\ 
103 &4.32$^f$&$4.22^{+0.26}_{-3.4}$& 1.7& 27.73 \\ 
104 &4.32$^f$&$3.54^{+0.86}_{-0.09}$&1.6 & 28.32 \\ 
105 &4.32$^f$&$4.20^{+0.11}_{-0.09}$&1.7 & 25.32\\ 
106 &4.32$^f$&$4.20^{+0.64}_{-0.64}$& 2.5 & 29.01\\ 
28a &4.32$^f$&$4.34^{+0.12}_{-0.12}$& 6.6 & 27.72\\ 
107 &4.32$^f$&$5.65^{+0.21}_{-4.7}$ & 11 & 28.91 \\ 
108 &4.32$^f$&$4.20^{+0.13}_{-0.14}$& 2.2 & 26.41\\ 
109 &4.32$^f$&$3.97^{+0.52}_{-0.38}$& 2.2 & 28.43\\ 
110 &4.32$^f$&$4.85^{+0.05}_{-1.4}$ & 3.6 & 29.39\\ 
111 &4.32$^f$&$4.22^{+0.09}_{-0.14}$& 1.6 & 25.32 \\ 
112 &4.32$^f$&$4.60^{+0.21}_{-3.7}$ & 2.2 & 26.96 \\ 
\enddata
\tablecomments{ Column 1: Image system ID; Column 2: redshift; Column 3: photometric redshift estimate from LePhare; Column 4: lensing magnification factor estimated from our lens model; Column 5: F200W apparent magnitude corrected for lensing magnification (``intrinsic'').}
\tablenotetext{a}{\citet{Caputi2021} measured a spectroscopic redshift at this coordinate.}
\tablenotetext{b}{A spectroscopic redshift was measured along the same arc and near  this specific coordinate in projection.}
\tablenotetext{c}{A spectroscopic redshift was measured for a counterimage of this arc system.}
\tablenotetext{d}{Lens-model redshift }
\tablenotetext{e}{MUSE single line detection, this study. } 
\tablenotetext{f}{Pure photometric selection, this study, with the redshift assumed}

  \label{tab_1}
\end{deluxetable}

To investigate this somewhat rare, strongly-lensed $z=4.32$ galaxy overdensity, it is beneficial to obtain observations longward of the 4000\,\AA\ and Balmer breaks (observed $\lambda\ga2$\,$\mu$m) to detect the redder stellar population and search for other members.
The new JWST PEARLS data provide rest-frame visible imaging to complement the rest-frame ultraviolet imaging from HST\null.   We  searched for additional lensed sources with similar redshifts predicted by the gravitational lens model ($z_{\rm mod}$) and for galaxies with similar photometric redshift estimates ($z_{\rm ph}$). The search identified one new probable member (system 39, Table~\ref{tab_1}). Figure~\ref{fig_z4p3} shows the known and one new probable group members. 

The high sensitivity and resolution of NIRCam
confirms image systems 8, 9, 10, and 11 by their similar morphologies, image orientations, and lens model predicted locations. In some cases, the higher resolution of NIRCam separates the images into additional components, strengthening the claim that galaxy associations and/or interactions may be instigating the star formation. The most striking example is system~9, the galaxy whose ``checkerboard" dual-color sides are detected in each of three images of this one galaxy.  The counterimages make obvious parity flips between images a and b and then again between images b and c, as predicted by lensing theory on each crossing of the critical curve. This parity information provides additional constraints on the lens model.  It is interesting to ask if these four galaxies at $z=4.32$ comprise of the entire galaxy group, or if they might be a subset of a larger galaxy overdensity.

\subsection{Search for new members} \label{sec:odense2}

We searched the \citet{Caminha2022} redshift catalog for other lensed sources that have velocity separations within $\pm 2800$~km~s$^{-1}$ of the systemic redshift of the initial group of galaxies, equating to $4.275 < z < 4.375$ \citep{Sifon2013,Caputi2021,Caminha2022}.  By these criteria, three other galaxies are identified, but only one of them is within the NIRCam field-of-view.  That one galaxy has a redshift of 4.2750 and a quality flag rating of `9' from \citet{Caminha2022}, which means that the emission feature is narrow or noisy with no secure identification and therefore fails to meet our criteria for a secure redshift.  We then searched the MUSE data cube for any emission lines which are consistent with being Ly$\alpha$ at $z=4.32$, an exercise that uncovered two spectroscopic candidates. Of these, only one lensed source (Object 101) survives our conservative requirements that the source must also be detected by NIRCam. Object 101 has  $z_{\rm sp}=4.317$ if the line is indeed Ly$\alpha$. The object is only marginally detected in NIRCam F200W, and its counterimages cannot be securely detected. The lens model gives a magnification factor of 2.1.



In addition to the blind spectroscopic search, we also searched for $z\approx4.32$ candidates based on redshifts predicted from our lens model ($z_{\rm  mod}$) and from our photometry ($z_{\rm  ph}$). To be considered a candidate, we required  that the lensed images have both $z_{\rm  mod}$ and $z_{\rm ph}$ consistent with $z=4.32$. One galaxy satisfied these criteria, image system 39, with $z_{\rm mod} = 4.14$. Arc 39a has $z_{\rm ph}=4.29 \pm 0.16$. At the system 39 positions, the MUSE data cube reveals only noise with no continuum or emission-line features anywhere in the spectrum covering 4700--9350~\AA.  However, at $z=4.32$, the only strong line in the passband is Ly$\alpha$, whose emission may be weak or even absent depending on the outflow velocity of the expanding \ion{H}{2} regions within which the line is produced \citep{Frye2002}.
The properties of these five members appear in Tables~\ref{tab_1} and \ref{tab_2}. For reference, the four confirmed members and system 39 have a mean intrinsic F200W  magnitude corrected for lensing magnificationof 25.75. This value is fainter than $L^*$ at F200W, which is $\approx$25.0 mag at $z = 4$ (R.~Cabanac, {private communication, 2023}).  

\begin{figure}[h]
\centering\includegraphics[scale=0.214]{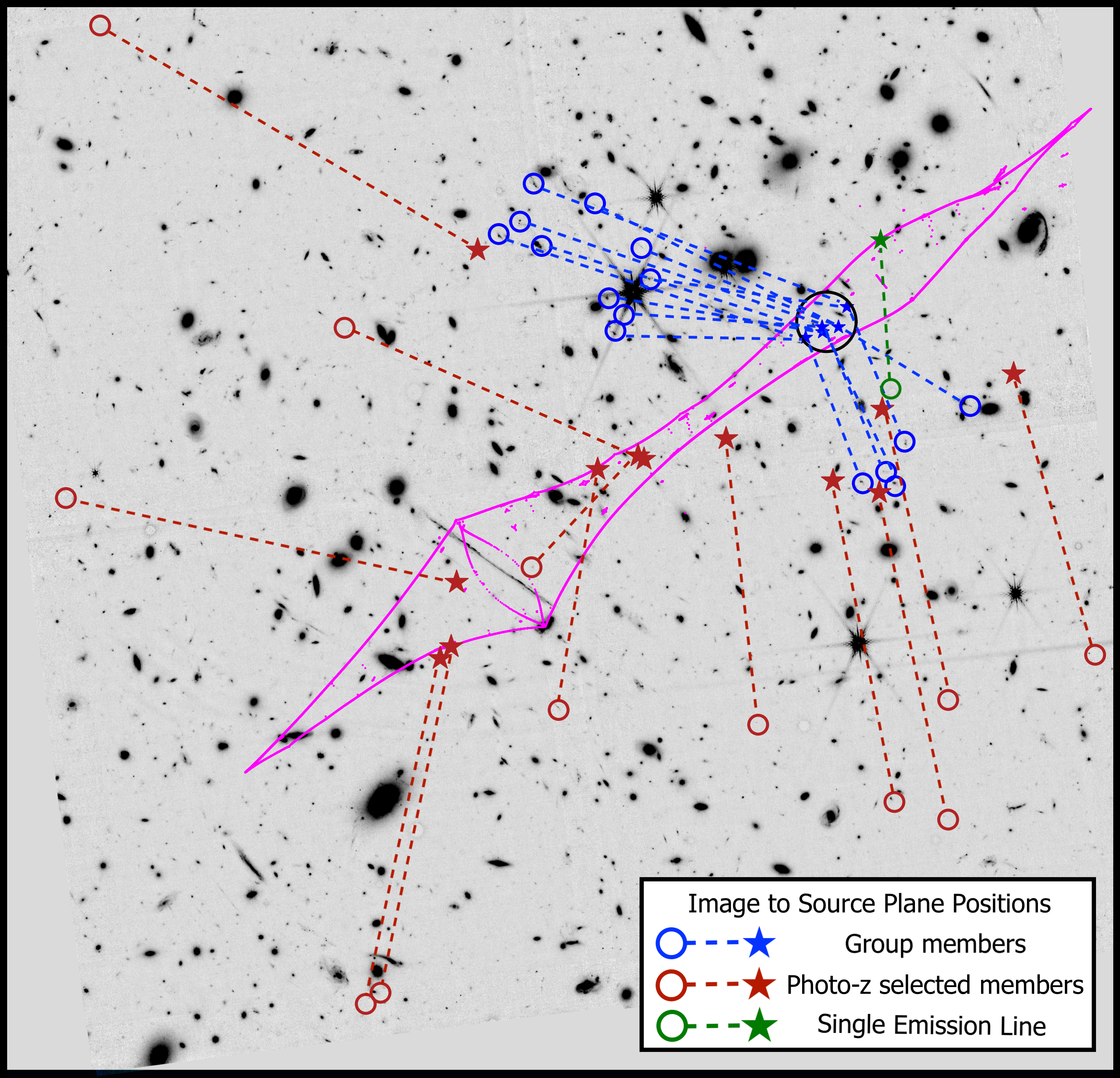}
\caption{NIRCam F200W negative image depicting the $z=4.32$ probable and candidate members. The image positions and source positions are designated by circular and stellar shapes, respectively.  The four galaxies at $z=4.32$ and one probable additional member with similar $z_{\rm mod}$ and $z_{\rm ph}$ values are lensed into 15 images (blue open circles)  which map onto the source plane as indicated by the blue stars. (These are image systems 8, 9, 10, 11, and 39 in Tables 1 and 2.) A circle with a radius of 30~kpc is overlayed for reference. The one galaxy obtained by a single-line detection in the MUSE data cube  is indicated in green. The 12 galaxies uncovered by a pure photometric search are indicated by the red open circles, which map onto the source plane at positions given by the red stars. These arcs correspond to systems 102--112 in Tables 1 and 3. The caustic for our LTM model is indicated by the solid pink contour.}
\label{fig_sourceplane}
\end{figure}

Lastly, we extended search for group members by a pure photometric approach, resulting in another 12 nonredundant lensed sources with $z_{\rm ph}=4.3$ within the 1$\sigma$ uncertainties. Positions of these objects are in Table~\ref{tab_3}.  Figure~\ref{fig_sourceplane} shows the source-plane positions of all 18 
sources. The five more secure members of the galaxy overdensity have a total physical extent of $\sim$60~kpc. The single-line spectroscopic detection, and the 12 photometrically-selected galaxies are more broadly distributed along the caustic and just outside it.  The caustic is elongated along the long axis of the cluster, and perhaps not surprisingly, the  group member candidates are relatively bright as a result of their high magnification factors of 1.5--11. The 12 photometrically-selected members have a mean apparent F200W magnitude corrected for lensing magnification that is 1.5 magnitudes fainter than the value for the five confirmed and probable members, pushing the limit of detectability in the HST filters. The deflections, as depicted by the dashed-line segments in Figure~\ref{fig_sourceplane}, do not vary much in angular extent.  This is consistent with lensing theory, which predicts that for the idealized case of an isothermal lens all sources at a similar source distance which are situated well behind the lens will be deflected by a constant amount \citep{Bartelmann1996}.

High-redshift galaxy overdensities can comprise of only a few galaxies, or in the most extreme case they can be sprawling structures extending up to several co-moving Mpc \citep{Chiang2017}. The fact that the five secure members of this galaxy overdensity are relatively bright and tightly-arranged, and that an additional 13 lensed sources are plausibly also at $z = 4.3$, suggests that this galaxy overdensity may be larger than just a compact group of $\sim$few galaxies. Four of the five secure members are also confirmed to be star-forming \citep{Caputi2021}, resolved (intrinsic physical extents of $\sim$1kpc), and  have morphologies indicative of galaxy interactions (Figure~\ref{fig_z4p3}).  By contrast, the 13 other candidate members are faint and small.  It is tempting to speculate whether these candidate members, if real, may have had their star formation rates truncated below 10~\% of their peak star forming episode, one definition of a quiescent galaxy \citep{Nanayakkara2023}.  Numerous NIRCam imaging studies have informed us that quiescent galaxies are more common than expected at $z=3-4$, implying an early and relatively-rapid buildup of stellar material in galaxies, and/or an efficient conversion rate of gas to stars at early times \citep{Nanayakkara2023, Carnall2023,  Valentino2023}.   A comprehensive investigation of the physical properties and star formation histories of these galaxies/candidates should uncover hints regarding their assembly, and their connection to any larger galaxy overdensity.

\section{Conclusions} \label{sec:finis}

A total of 56 multiply-imaged galaxies are vetted by our LTM model, including two new image systems and five new counterimages, showcasing the benefits of the high-sensitivity and high-resolution NIRCam imaging. The incorporation of HST imaging enables PSF-corrected photometry across the visible and NIR passbands for all sources in the region of overlap of the 11 bands.  The mass estimated from the lens model within 500~kpc is $(7.0 \pm 0.30) \times 10^{14}$~\Msol\ with a mass ratio between the SE and NW components close to unity.   The SE peak is centered near the BCG, and the X-ray peak is centered very near the cooling flow of the BCG, in agreement with recent studies. 

Although the two mass peaks are 650~kpc (projected distance) apart in the imaging, they are close in velocity space.  This suggests a significant transverse velocity component, and in fact a statistical search is required to uncover the SE and NW components.  The dynamical mass measured from the galaxy velocity dispersion in a double-halo configuration is $(5.1\pm0.60)\times10^{14}$~\Msol.  This value is based on 51 more galaxies than a previous estimate to similar radius.  The measured mass is lower than the value estimated by strong- and weak-lensing studies.  This difference can be explained if the transverse velocity is high, therefore underestimating the dynamical mass. 

A motivation of this study was the search for additional galaxy members in the previously-known $z=4.32$ galaxy overdensity. This new study was made feasible by the high sensitivity of NIRCam, enabling the detection of $z=4.32$ galaxies by extending the wavelength reach longer than the 4000~\AA\ and Balmer breaks, and at depths equating to $\sim$\,few magnitudes below M$^*$. 
By the combination of our model-predicted and photometric redshift estimates, one new member was discovered with high probability, and another possible member was discovered by a single emission line, assumed to be Ly$\alpha$ in the MUSE data cube. A pure photometric-redshift search identified another dozen candidates. If real, then the addition of these 14 galaxies more than quadruples the total number of galaxies in this rare view of a strongly-lensed structure. Given the broad range in apparent magnitudes, it tempting to ask if there may be an even more diverse population of galaxies underlying the confirmed members, some of which may be already be quiescent. Ultimately, spectroscopic confirmation of these candidate members is needed to establish the nature of this rare view of a strongly-lensed galaxy overdensity.

\acknowledgments 
We dedicate this study to the memory of Jill Bechtold, scholar and mentor, who with her great patience and investment in undergraduate and graduate education set many of us onto a career path in astronomy.
We thank Sergey Cherkis for useful conversations and the anonymous referee for suggestions that improved the manuscript.
B.L.F.~obtained student support through a Faculty Challenge Grant for Increasing Access to Undergraduate Research, and the Arthur L. and Lee G. Herbst Endowment for Innovation and the Science Dean’s Innovation and Education Fund, both obtained at the University of Arizona. R.A.W.~was funded by NASA JWST Interdisciplinary Scientist grants NAG5-12460, NNX14AN10G, and 80GNSSC18K0200 from NASA Goddard Space Flight Center. The BGU lensing group, LJF and AZ, acknowledge support by grant 2020750 from the United States-Israel Bi-national Science Foundation (BSF) and grant 2109066 from the United States National Science Foundation (NSF), and by the Ministry of Science \& Technology, Israel. KIC acknowledges funding from the Netherlands Research School for Astronomy (NOVA), and also from the Dutch Research Council (NWO) through the award of the Vici Grant VI.C.212.036.
We thank the JWST Project at NASA GSFC and JWST Program at NASA HQ for their many-decades long dedication to make the JWST mission a success. We especially thank Tony Roman, the JWST scheduling group and Mission Operations Center staff at STScI for their continued dedicated support to get the JWST observations scheduled. This work is based on observations made with the NASA/ESA/CSA James Webb Space Telescope. The data were obtained from the Mikulski Archive for Space Telescopes (MAST) at the Space Telescope Science Institute, which is operated by the Associa- tion of Universities for Research in Astronomy, Inc., under NASA contract NAS 5-03127 for JWST. These observations are associated with JWST programs 1176. This work is also based on observations made with the NASA/ESA {\it Hubble Space Telescope} (HST).  The data were obtained from the {\tt Barbara A.~Mikulski Archive for Space Telescopes (MAST)} at the {\it Space Telescope Science Institute} (STScI), which is operated by the Association of Universities for Research in Astronomy (AURA) Inc., under NASA contract NAS 5-26555 for HST.  This research has made use of data obtained from the {\it Chandra} Data Archive and software provided by the {\it Chandra} X-ray Center (CXC) in the application package CIAO.  


\begin{thebibliography}{}
\expandafter\ifx\csname natexlab\endcsname\relax\def\natexlab#1{#1}\fi
\providecommand{\url}[1]{\href{#1}{#1}}
\providecommand{\dodoi}[1]{doi:~\href{http://doi.org/#1}{\nolinkurl{#1}}}
\providecommand{\doeprint}[1]{\href{http://ascl.net/#1}{\nolinkurl{http://ascl.net/#1}}}
\providecommand{\doarXiv}[1]{\href{https://arxiv.org/abs/#1}{\nolinkurl{https://arxiv.org/abs/#1}}}

\bibitem[{{Arnouts} \& {Ilbert}(2011)}]{Arnouts2011}
{Arnouts}, S., \& {Ilbert}, O. 2011, {LePHARE: Photometric Analysis for
  Redshift Estimate}, Astrophysics Source Code Library, record ascl:1108.009.
\newblock \doeprint{1108.009}

\bibitem[{{Beers} {et~al.}(1990){Beers}, {Flynn}, \& {Gebhardt}}]{Beers1990}
{Beers}, T.~C., {Flynn}, K., \& {Gebhardt}, K. 1990, \aj, 100, 32,
  \dodoi{10.1086/115487}

\bibitem[{{Bertin} \& {Arnouts}(1996)}]{Bertin1996}
{Bertin}, E., \& {Arnouts}, S. 1996, \aaps, 117, 393,
  \dodoi{10.1051/aas:1996164}

\bibitem[{{Biviano} {et~al.}(2002){Biviano}, {Katgert}, {Thomas}, \&
  {Adami}}]{Biviano2002}
{Biviano}, A., {Katgert}, P., {Thomas}, T., \& {Adami}, C. 2002, \aap, 387, 8,
  \dodoi{10.1051/0004-6361:20020340}

\bibitem[{{Bradley} {et~al.}(2022){Bradley}, {Sip{\H{o}}cz}, {Robitaille},
  {Tollerud}, {Vin{\'\i}cius}, {Deil}, {Barbary}, {Wilson}, {Busko}, {Donath},
  {G{\"u}nther}, {Cara}, {Lim}, {Me{\ss}linger}, {Conseil}, {Bostroem},
  {Droettboom}, {Bray}, {Andersen Bratholm}, {Barentsen}, {Craig}, {Ginsburg},
  {Rathi}, {Pascual}, {Perren}, {Georgiev}, {De Val-Borro}, {Kerzendorf},
  {Bach}, \& {Quint}}]{Bradley2022}
{Bradley}, L., {Sip{\H{o}}cz}, B., {Robitaille}, T., {et~al.} 2022,
  {astropy/photutils: 1.6.0}, 1.6.0, Zenodo,  Zenodo,
  \dodoi{10.5281/zenodo.7419741}

\bibitem[{Brammer(2021)}]{Brammer2021}
Brammer, G. 2021, {eazy-py}, 0.5.2, \dodoi{10.5281/zenodo.5012704}

\bibitem[{{Brammer} {et~al.}(2008){Brammer}, {van Dokkum}, \&
  {Coppi}}]{Brammer2008}
{Brammer}, G.~B., {van Dokkum}, P.~G., \& {Coppi}, P. 2008, \apj, 686, 1503,
  \dodoi{10.1086/591786}

\bibitem[{{Caminha} {et~al.}(2022){Caminha}, {Grillo}, {Rosati}, {Liu},
  {Acebron}, {Bergamini}, {Caputi}, {Mercurio}, {Tozzi}, {Vanzella}, {Demarco},
  {Frye}, {Rosani}, \& {Sharon}}]{Caminha2022}
{Caminha}, G.~B., {Grillo}, C., {Rosati}, P., {et~al.} 2022, arXiv e-prints,
  arXiv:2209.02718.
\newblock \doarXiv{2209.02718}

\bibitem[{{Caputi} {et~al.}(2021){Caputi}, {Caminha}, {Fujimoto}, {Kohno},
  {Sun}, {Egami}, {Deshmukh}, {Tang}, {Ao}, {Bradley}, {Coe}, {Espada},
  {Grillo}, {Hatsukade}, {Knudsen}, {Lee}, {Magdis}, {Morokuma-Matsui},
  {Oesch}, {Ouchi}, {Rosati}, {Umehata}, {Valentino}, {Vanzella}, {Wang}, {Wu},
  \& {Zitrin}}]{Caputi2021}
{Caputi}, K.~I., {Caminha}, G.~B., {Fujimoto}, S., {et~al.} 2021, \apj, 908,
  146, \dodoi{10.3847/1538-4357/abd4d0}

\bibitem[{{Carnall} {et~al.}(2023){Carnall}, {McLeod}, {McLure}, {Dunlop},
  {Begley}, {Cullen}, {Donnan}, {Hamadouche}, {Jewell}, {Jones}, {Pollock}, \&
  {Wild}}]{Carnall2023}
{Carnall}, A.~C., {McLeod}, D.~J., {McLure}, R.~J., {et~al.} 2023, \mnras,
  \dodoi{10.1093/mnras/stad369}

\bibitem[{{Cerny} {et~al.}(2018){Cerny}, {Sharon}, {Andrade-Santos}, {Avila},
  {Brada{\v{c}}}, {Bradley}, {Carrasco}, {Coe}, {Czakon}, {Dawson}, {Frye},
  {Hoag}, {Huang}, {Johnson}, {Jones}, {Lam}, {Lovisari}, {Mainali}, {Oesch},
  {Ogaz}, {Past}, {Paterno-Mahler}, {Peterson}, {Riess}, {Rodney}, {Ryan},
  {Salmon}, {Sendra-Server}, {Stark}, {Strolger}, {Trenti}, {Umetsu},
  {Vulcani}, \& {Zitrin}}]{Cerny2018}
{Cerny}, C., {Sharon}, K., {Andrade-Santos}, F., {et~al.} 2018, \apj, 859, 159,
  \dodoi{10.3847/1538-4357/aabe7b}

\bibitem[{{Chiang} {et~al.}(2017){Chiang}, {Overzier}, {Gebhardt}, \&
  {Henriques}}]{Chiang2017}
{Chiang}, Y.-K., {Overzier}, R.~A., {Gebhardt}, K., \& {Henriques}, B. 2017,
  \apjl, 844, L23, \dodoi{10.3847/2041-8213/aa7e7b}

\bibitem[{{Coe} {et~al.}(2006){Coe}, {Ben{\'\i}tez}, {S{\'a}nchez}, {Jee},
  {Bouwens}, \& {Ford}}]{Coe2006}
{Coe}, D., {Ben{\'\i}tez}, N., {S{\'a}nchez}, S.~F., {et~al.} 2006, \aj, 132,
  926, \dodoi{10.1086/505530}

\bibitem[{{Coe} {et~al.}(2012){Coe}, {Umetsu}, {Zitrin}, {Donahue},
  {Medezinski}, {Postman}, {Carrasco}, {Anguita}, {Geller}, {Rines},
  {Diaferio}, {Kurtz}, {Bradley}, {Koekemoer}, {Zheng}, {Nonino}, {Molino},
  {Mahdavi}, {Lemze}, {Infante}, {Ogaz}, {Melchior}, {Host}, {Ford}, {Grillo},
  {Rosati}, {Jim{\'e}nez-Teja}, {Moustakas}, {Broadhurst}, {Ascaso}, {Lahav},
  {Bartelmann}, {Ben{\'\i}tez}, {Bouwens}, {Graur}, {Graves}, {Jha}, {Jouvel},
  {Kelson}, {Moustakas}, {Maoz}, {Meneghetti}, {Merten}, {Riess}, {Rodney}, \&
  {Seitz}}]{Coe2012}
{Coe}, D., {Umetsu}, K., {Zitrin}, A., {et~al.} 2012, \apj, 757, 22,
  \dodoi{10.1088/0004-637X/757/1/22}

\bibitem[{{Diego} {et~al.}(2020){Diego}, {Molnar}, {Cerny}, {Broadhurst},
  {Windhorst}, {Zitrin}, {Bouwens}, {Coe}, {Conselice}, \&
  {Sharon}}]{Diego2020}
{Diego}, J.~M., {Molnar}, S.~M., {Cerny}, C., {et~al.} 2020, \apj, 904, 106,
  \dodoi{10.3847/1538-4357/abbf56}

\bibitem[{{Diego} {et~al.}(2023){Diego}, {Meena}, {Adams}, {Broadhurst}, {Dai},
  {Coe}, {Frye}, {Kelly}, {Koekemoer}, {Pascale}, {Willner}, {Zackrisson},
  {Zitrin}, {Windhorst}, {Cohen}, {Jansen}, {Summers}, {Tompkins}, {Conselice},
  {Driver}, {Yan}, {Grogin}, {Marshall}, {Pirzkal}, {Robotham}, {Ryan},
  {Willmer}, {Bradley}, {Caminha}, {Caputi}, {Carleton}, \&
  {Kamieneski}}]{Diego2023}
{Diego}, J.~M., {Meena}, A.~K., {Adams}, N.~J., {et~al.} 2023, \aap, 672, A3,
  \dodoi{10.1051/0004-6361/202245238}

\bibitem[{{Dressler} \& {Shectman}(1988)}]{Dressler1988}
{Dressler}, A., \& {Shectman}, S.~A. 1988, \aj, 95, 985, \dodoi{10.1086/114694}

\bibitem[{{Fruscione} {et~al.}(2006){Fruscione}, {McDowell}, {Allen},
  {Brickhouse}, {Burke}, {Davis}, {Durham}, {Elvis}, {Galle}, {Harris},
  {Huenemoerder}, {Houck}, {Ishibashi}, {Karovska}, {Nicastro}, {Noble},
  {Nowak}, {Primini}, {Siemiginowska}, {Smith}, \& {Wise}}]{Fruscione2006}
{Fruscione}, A., {McDowell}, J.~C., {Allen}, G.~E., {et~al.} 2006, in Society
  of Photo-Optical Instrumentation Engineers (SPIE) Conference Series, Vol.
  6270, Society of Photo-Optical Instrumentation Engineers (SPIE) Conference
  Series, ed. D.~R. {Silva} \& R.~E. {Doxsey}, 62701V,
  \dodoi{10.1117/12.671760}

\bibitem[{{Frye} {et~al.}(2002){Frye}, {Broadhurst}, \&
  {Ben{\'\i}tez}}]{Frye2002}
{Frye}, B., {Broadhurst}, T., \& {Ben{\'\i}tez}, N. 2002, \apj, 568, 558,
  \dodoi{10.1086/338965}

\bibitem[{{Galametz} {et~al.}(2013){Galametz}, {Grazian}, {Fontana},
  {Ferguson}, {Ashby}, {Barro}, {Castellano}, {Dahlen}, {Donley}, {Faber},
  {Grogin}, {Guo}, {Huang}, {Kocevski}, {Koekemoer}, {Lee}, {McGrath}, {Peth},
  {Willner}, {Almaini}, {Cooper}, {Cooray}, {Conselice}, {Dickinson}, {Dunlop},
  {Fazio}, {Foucaud}, {Gardner}, {Giavalisco}, {Hathi}, {Hartley}, {Koo},
  {Lai}, {de Mello}, {McLure}, {Lucas}, {Paris}, {Pentericci}, {Santini},
  {Simpson}, {Sommariva}, {Targett}, {Weiner}, {Wuyts}, \& {CANDELS
  Team}}]{Galametz2013}
{Galametz}, A., {Grazian}, A., {Fontana}, A., {et~al.} 2013, \apjs, 206, 10,
  \dodoi{10.1088/0067-0049/206/2/10}

\bibitem[{{Jee} {et~al.}(2014){Jee}, {Hughes}, {Menanteau}, {Sif{\'o}n},
  {Mandelbaum}, {Barrientos}, {Infante}, \& {Ng}}]{Jee2014}
{Jee}, M.~J., {Hughes}, J.~P., {Menanteau}, F., {et~al.} 2014, \apj, 785, 20,
  \dodoi{10.1088/0004-637X/785/1/20}

\bibitem[{{Johnson} \& {Sharon}(2016)}]{Johnson2016}
{Johnson}, T.~L., \& {Sharon}, K. 2016, \apj, 832, 82,
  \dodoi{10.3847/0004-637X/832/1/82}

\bibitem[{{Kamieneski} {et~al.}(2023){Kamieneski}, {Frye}, {Pascale}, {Cohen},
  {Windhorst}, {Jansen}, {Yun}, {Cheng}, {Summers}, {Carleton}, {Harrington},
  {Diego}, {Yan}, {Koekemoer}, {Willmer}, {Petric}, {Furtak}, {Foo},
  {Conselice}, {Coe}, {Driver}, {Grogin}, {Marshall}, {Pirzkal}, {Robotham},
  {Ryan}, \& {Tompkins}}]{Kamieneski2023}
{Kamieneski}, P.~S., {Frye}, B.~L., {Pascale}, M., {et~al.} 2023, arXiv
  e-prints, arXiv:2303.05054, \dodoi{10.48550/arXiv.2303.05054}

\bibitem[{{Kim} {et~al.}(2021){Kim}, {Jee}, {Hughes}, {Yoon}, {HyeongHan},
  {Menanteau}, {Sif{\'o}n}, {Hovey}, \& {Arunachalam}}]{Kim2021}
{Kim}, J., {Jee}, M.~J., {Hughes}, J.~P., {et~al.} 2021, \apj, 923, 101,
  \dodoi{10.3847/1538-4357/ac294f}

\bibitem[{{Koekemoer} {et~al.}(2011){Koekemoer}, {Faber}, {Ferguson}, {Grogin},
  {Kocevski}, {Koo}, {Lai}, {Lotz}, {Lucas}, {McGrath}, {Ogaz}, {Rajan},
  {Riess}, {Rodney}, {Strolger}, {Casertano}, {Castellano}, {Dahlen},
  {Dickinson}, {Dolch}, {Fontana}, {Giavalisco}, {Grazian}, {Guo}, {Hathi},
  {Huang}, {van der Wel}, {Yan}, {Acquaviva}, {Alexander}, {Almaini}, {Ashby},
  {Barden}, {Bell}, {Bournaud}, {Brown}, {Caputi}, {Cassata}, {Challis},
  {Chary}, {Cheung}, {Cirasuolo}, {Conselice}, {Roshan Cooray}, {Croton},
  {Daddi}, {Dav{\'e}}, {de Mello}, {de Ravel}, {Dekel}, {Donley}, {Dunlop},
  {Dutton}, {Elbaz}, {Fazio}, {Filippenko}, {Finkelstein}, {Frazer}, {Gardner},
  {Garnavich}, {Gawiser}, {Gruetzbauch}, {Hartley}, {H{\"a}ussler},
  {Herrington}, {Hopkins}, {Huang}, {Jha}, {Johnson}, {Kartaltepe},
  {Khostovan}, {Kirshner}, {Lani}, {Lee}, {Li}, {Madau}, {McCarthy},
  {McIntosh}, {McLure}, {McPartland}, {Mobasher}, {Moreira}, {Mortlock},
  {Moustakas}, {Mozena}, {Nandra}, {Newman}, {Nielsen}, {Niemi}, {Noeske},
  {Papovich}, {Pentericci}, {Pope}, {Primack}, {Ravindranath}, {Reddy},
  {Renzini}, {Rix}, {Robaina}, {Rosario}, {Rosati}, {Salimbeni}, {Scarlata},
  {Siana}, {Simard}, {Smidt}, {Snyder}, {Somerville}, {Spinrad}, {Straughn},
  {Telford}, {Teplitz}, {Trump}, {Vargas}, {Villforth}, {Wagner}, {Wandro},
  {Wechsler}, {Weiner}, {Wiklind}, {Wild}, {Wilson}, {Wuyts}, \&
  {Yun}}]{Koekemoer2011}
{Koekemoer}, A.~M., {Faber}, S.~M., {Ferguson}, H.~C., {et~al.} 2011, \apjs,
  197, 36, \dodoi{10.1088/0067-0049/197/2/36}

\bibitem[{{Krist} {et~al.}(2011){Krist}, {Hook}, \& {Stoehr}}]{Krist2011}
{Krist}, J.~E., {Hook}, R.~N., \& {Stoehr}, F. 2011, in Society of
  Photo-Optical Instrumentation Engineers (SPIE) Conference Series, Vol. 8127,
  Optical Modeling and Performance Predictions V, ed. M.~A. {Kahan}, 81270J,
  \dodoi{10.1117/12.892762}

\bibitem[{{Larson} {et~al.}(2022){Larson}, {Hutchison}, {Bagley},
  {Finkelstein}, {Yung}, {Somerville}, {Hirschmann}, {Brammer}, {Holwerda},
  {Papovich}, {Morales}, \& {Wilkins}}]{Larson2022}
{Larson}, R.~L., {Hutchison}, T.~A., {Bagley}, M., {et~al.} 2022, arXiv
  e-prints, arXiv:2211.10035.
\newblock \doarXiv{2211.10035}

\bibitem[{{Lindner} {et~al.}(2014){Lindner}, {Baker}, {Hughes}, {Battaglia},
  {Gupta}, {Knowles}, {Marriage}, {Menanteau}, {Moodley}, {Reese}, \&
  {Srianand}}]{Linder2014}
{Lindner}, R.~R., {Baker}, A.~J., {Hughes}, J.~P., {et~al.} 2014, \apj, 786,
  49, \dodoi{10.1088/0004-637X/786/1/49}

\bibitem[{{Marriage} {et~al.}(2011){Marriage}, {Acquaviva}, {Ade}, {Aguirre},
  {Amiri}, {Appel}, {Barrientos}, {Battistelli}, {Bond}, {Brown}, {Burger},
  {Chervenak}, {Das}, {Devlin}, {Dicker}, {Bertrand Doriese}, {Dunkley},
  {D{\"u}nner}, {Essinger-Hileman}, {Fisher}, {Fowler}, {Hajian}, {Halpern},
  {Hasselfield}, {Hern{\'a}ndez-Monteagudo}, {Hilton}, {Hilton}, {Hincks},
  {Hlozek}, {Huffenberger}, {Handel Hughes}, {Hughes}, {Infante}, {Irwin},
  {Baptiste Juin}, {Kaul}, {Klein}, {Kosowsky}, {Lau}, {Limon}, {Lin},
  {Lupton}, {Marsden}, {Martocci}, {Mauskopf}, {Menanteau}, {Moodley},
  {Moseley}, {Netterfield}, {Niemack}, {Nolta}, {Page}, {Parker}, {Partridge},
  {Quintana}, {Reese}, {Reid}, {Sehgal}, {Sherwin}, {Sievers}, {Spergel},
  {Staggs}, {Swetz}, {Switzer}, {Thornton}, {Trac}, {Tucker}, {Warne},
  {Wilson}, {Wollack}, \& {Zhao}}]{Marriage2011}
{Marriage}, T.~A., {Acquaviva}, V., {Ade}, P. A.~R., {et~al.} 2011, \apj, 737,
  61, \dodoi{10.1088/0004-637X/737/2/61}

\bibitem[{{McKinney} {et~al.}(2022){McKinney}, {Finnerty}, {Casey}, {Franco},
  {Long}, {Fujimoto}, {Zavala}, {Cooper}, {Akins}, {Pope}, {Armus}, {Soifer},
  {Larson}, {Matthews}, {Melbourne}, \& {Cushing}}]{McKinney2023}
{McKinney}, J., {Finnerty}, L., {Casey}, C., {et~al.} 2022, arXiv e-prints,
  arXiv:2301.00017, \dodoi{10.48550/arXiv.2301.00017}

\bibitem[{{Menanteau} {et~al.}(2012){Menanteau}, {Hughes}, {Sif{\'o}n},
  {Hilton}, {Gonz{\'a}lez}, {Infante}, {Barrientos}, {Baker}, {Bond}, {Das},
  {Devlin}, {Dunkley}, {Hajian}, {Hincks}, {Kosowsky}, {Marsden}, {Marriage},
  {Moodley}, {Niemack}, {Nolta}, {Page}, {Reese}, {Sehgal}, {Sievers},
  {Spergel}, {Staggs}, \& {Wollack}}]{Menanteau2012}
{Menanteau}, F., {Hughes}, J.~P., {Sif{\'o}n}, C., {et~al.} 2012, \apj, 748, 7,
  \dodoi{10.1088/0004-637X/748/1/7}

\bibitem[{{Meneghetti} {et~al.}(2017){Meneghetti}, {Natarajan}, {Coe},
  {Contini}, {De Lucia}, {Giocoli}, {Acebron}, {Borgani}, {Bradac}, {Diego},
  {Hoag}, {Ishigaki}, {Johnson}, {Jullo}, {Kawamata}, {Lam}, {Limousin},
  {Liesenborgs}, {Oguri}, {Sebesta}, {Sharon}, {Williams}, \&
  {Zitrin}}]{Meneghetti2017}
{Meneghetti}, M., {Natarajan}, P., {Coe}, D., {et~al.} 2017, \mnras, 472, 3177,
  \dodoi{10.1093/mnras/stx2064}

\bibitem[{{Molnar} \& {Broadhurst}(2015)}]{Molnar2015}
{Molnar}, S.~M., \& {Broadhurst}, T. 2015, \apj, 800, 37,
  \dodoi{10.1088/0004-637X/800/1/37}

\bibitem[{{Mortonson} {et~al.}(2011){Mortonson}, {Hu}, \&
  {Huterer}}]{Mortonson2011}
{Mortonson}, M.~J., {Hu}, W., \& {Huterer}, D. 2011, \prd, 83, 023015,
  \dodoi{10.1103/PhysRevD.83.023015}

\bibitem[{{Nanayakkara} {et~al.}(2022){Nanayakkara}, {Glazebrook}, {Jacobs},
  {Schreiber}, {Brammer}, {Esdaile}, {Kacprzak}, {Labbe}, {Lagos},
  {Marchesini}, {Marsan}, {Nateghi}, {Oesch}, {Papovich}, {Remus}, \&
  {Tran}}]{Nanayakkara2023}
{Nanayakkara}, T., {Glazebrook}, K., {Jacobs}, C., {et~al.} 2022, arXiv
  e-prints, arXiv:2212.11638, \dodoi{10.48550/arXiv.2212.11638}

\bibitem[{{Narayan} \& {Bartelmann}(1996)}]{Bartelmann1996}
{Narayan}, R., \& {Bartelmann}, M. 1996, arXiv e-prints, astro,
  \dodoi{10.48550/arXiv.astro-ph/9606001}

\bibitem[{{Ng} {et~al.}(2015){Ng}, {Dawson}, {Wittman}, {Jee}, {Hughes},
  {Menanteau}, \& {Sif{\'o}n}}]{Ng2015}
{Ng}, K.~Y., {Dawson}, W.~A., {Wittman}, D., {et~al.} 2015, \mnras, 453, 1531,
  \dodoi{10.1093/mnras/stv1713}

\bibitem[{{Pascale} {et~al.}(2022){Pascale}, {Frye}, {Dai}, {Foo}, {Qin},
  {Leimbach}, {Bauer}, {Merlin}, {Coe}, {Diego}, {Yan}, {Zitrin}, {Cohen},
  {Conselice}, {Dole}, {Harrington}, {Jansen}, {Kamieneski}, {Windhorst}, \&
  {Yun}}]{Pascale2022a}
{Pascale}, M., {Frye}, B.~L., {Dai}, L., {et~al.} 2022, \apj, 932, 85,
  \dodoi{10.3847/1538-4357/ac6ce9}

\bibitem[{{Planck Collaboration} {et~al.}(2020){Planck Collaboration},
  {Aghanim}, {Akrami}, {Ashdown}, {Aumont}, {Baccigalupi}, {Ballardini},
  {Banday}, {Barreiro}, {Bartolo}, {Basak}, {Battye}, {Benabed}, {Bernard},
  {Bersanelli}, {Bielewicz}, {Bock}, {Bond}, {Borrill}, {Bouchet}, {Boulanger},
  {Bucher}, {Burigana}, {Butler}, {Calabrese}, {Cardoso}, {Carron},
  {Challinor}, {Chiang}, {Chluba}, {Colombo}, {Combet}, {Contreras}, {Crill},
  {Cuttaia}, {de Bernardis}, {de Zotti}, {Delabrouille}, {Delouis}, {Di
  Valentino}, {Diego}, {Dor{\'e}}, {Douspis}, {Ducout}, {Dupac}, {Dusini},
  {Efstathiou}, {Elsner}, {En{\ss}lin}, {Eriksen}, {Fantaye}, {Farhang},
  {Fergusson}, {Fernandez-Cobos}, {Finelli}, {Forastieri}, {Frailis},
  {Fraisse}, {Franceschi}, {Frolov}, {Galeotta}, {Galli}, {Ganga},
  {G{\'e}nova-Santos}, {Gerbino}, {Ghosh}, {Gonz{\'a}lez-Nuevo}, {G{\'o}rski},
  {Gratton}, {Gruppuso}, {Gudmundsson}, {Hamann}, {Handley}, {Hansen},
  {Herranz}, {Hildebrandt}, {Hivon}, {Huang}, {Jaffe}, {Jones}, {Karakci},
  {Keih{\"a}nen}, {Keskitalo}, {Kiiveri}, {Kim}, {Kisner}, {Knox},
  {Krachmalnicoff}, {Kunz}, {Kurki-Suonio}, {Lagache}, {Lamarre}, {Lasenby},
  {Lattanzi}, {Lawrence}, {Le Jeune}, {Lemos}, {Lesgourgues}, {Levrier},
  {Lewis}, {Liguori}, {Lilje}, {Lilley}, {Lindholm}, {L{\'o}pez-Caniego},
  {Lubin}, {Ma}, {Mac{\'\i}as-P{\'e}rez}, {Maggio}, {Maino}, {Mandolesi},
  {Mangilli}, {Marcos-Caballero}, {Maris}, {Martin}, {Martinelli},
  {Mart{\'\i}nez-Gonz{\'a}lez}, {Matarrese}, {Mauri}, {McEwen}, {Meinhold},
  {Melchiorri}, {Mennella}, {Migliaccio}, {Millea}, {Mitra},
  {Miville-Desch{\^e}nes}, {Molinari}, {Montier}, {Morgante}, {Moss}, {Natoli},
  {N{\o}rgaard-Nielsen}, {Pagano}, {Paoletti}, {Partridge}, {Patanchon},
  {Peiris}, {Perrotta}, {Pettorino}, {Piacentini}, {Polastri}, {Polenta},
  {Puget}, {Rachen}, {Reinecke}, {Remazeilles}, {Renzi}, {Rocha}, {Rosset},
  {Roudier}, {Rubi{\~n}o-Mart{\'\i}n}, {Ruiz-Granados}, {Salvati}, {Sandri},
  {Savelainen}, {Scott}, {Shellard}, {Sirignano}, {Sirri}, {Spencer},
  {Sunyaev}, {Suur-Uski}, {Tauber}, {Tavagnacco}, {Tenti}, {Toffolatti},
  {Tomasi}, {Trombetti}, {Valenziano}, {Valiviita}, {Van Tent}, {Vibert},
  {Vielva}, {Villa}, {Vittorio}, {Wandelt}, {Wehus}, {White}, {White},
  {Zacchei}, \& {Zonca}}]{PlanckCollaboration2018}
{Planck Collaboration}, {Aghanim}, N., {Akrami}, Y., {et~al.} 2020, \aap, 641,
  A6, \dodoi{10.1051/0004-6361/201833910}

\bibitem[{{Rieke} {et~al.}(2005){Rieke}, {Kelly}, \& {Horner}}]{Rieke05}
{Rieke}, M.~J., {Kelly}, D., \& {Horner}, S. 2005, in Society of Photo-Optical
  Instrumentation Engineers (SPIE) Conference Series, Vol. 5904, Cryogenic
  Optical Systems and Instruments XI, ed. J.~B. {Heaney} \& L.~G. {Burriesci},
  1--8, \dodoi{10.1117/12.615554}

\bibitem[{{Robotham} {et~al.}(2018){Robotham}, {Davies}, {Driver}, {Koushan},
  {Taranud}, {Casura}, \& {Liske}}]{Robotham2018}
{Robotham}, A.~S.~G., {Davies}, L.~J.~M., {Driver}, S.~P., {et~al.} 2018,
  \mnras, 476, 3137, \dodoi{10.1093/mnras/sty440}

\bibitem[{{Robotham} {et~al.}(2017){Robotham}, {Taranu}, {Tobar}, {Moffett}, \&
  {Driver}}]{Robotham2017}
{Robotham}, A.~S.~G., {Taranu}, D.~S., {Tobar}, R., {Moffett}, A., \& {Driver},
  S.~P. 2017, \mnras, 466, 1513, \dodoi{10.1093/mnras/stw3039}

\bibitem[{{Rumbaugh} {et~al.}(2018){Rumbaugh}, {Lemaux}, {Tomczak}, {Shen},
  {Pelliccia}, {Lubin}, {Kocevski}, {Wu}, {Gal}, {Mei}, {Fassnacht}, \&
  {Squires}}]{Rumbaugh2018}
{Rumbaugh}, N., {Lemaux}, B.~C., {Tomczak}, A.~R., {et~al.} 2018, \mnras, 478,
  1403, \dodoi{10.1093/mnras/sty1181}

\bibitem[{{Sif{\'o}n} {et~al.}(2013){Sif{\'o}n}, {Menanteau}, {Hasselfield},
  {Marriage}, {Hughes}, {Barrientos}, {Gonz{\'a}lez}, {Infante}, {Addison},
  {Baker}, {Battaglia}, {Bond}, {Crichton}, {Das}, {Devlin}, {Dunkley},
  {D{\"u}nner}, {Gralla}, {Hajian}, {Hilton}, {Hincks}, {Kosowsky}, {Marsden},
  {Moodley}, {Niemack}, {Nolta}, {Page}, {Partridge}, {Reese}, {Sehgal},
  {Sievers}, {Spergel}, {Staggs}, {Thornton}, {Trac}, \& {Wollack}}]{Sifon2013}
{Sif{\'o}n}, C., {Menanteau}, F., {Hasselfield}, M., {et~al.} 2013, \apj, 772,
  25, \dodoi{10.1088/0004-637X/772/1/25}

\bibitem[{{Strait} {et~al.}(2018){Strait}, {Brada{\v{c}}}, {Hoag}, {Huang},
  {Treu}, {Wang}, {Amorin}, {Castellano}, {Fontana}, {Lemaux}, {Merlin},
  {Schmidt}, {Schrabback}, {Tomczack}, {Trenti}, \& {Vulcani}}]{Strait2018}
{Strait}, V., {Brada{\v{c}}}, M., {Hoag}, A., {et~al.} 2018, \apj, 868, 129,
  \dodoi{10.3847/1538-4357/aae834}

\bibitem[{{Valentino} {et~al.}(2023){Valentino}, {Brammer}, {Gould}, {Kokorev},
  {Fujimoto}, {Kragh Jespersen}, {Vijayan}, {Weaver}, {Ito}, {Tanaka},
  {Ilbert}, {Magdis}, {Whitaker}, {Faisst}, {Gallazzi}, {Gillman},
  {Gimenez-Arteaga}, {Gomez-Guijarro}, {Kubo}, {Heintz}, {Hirschmann}, {Oesch},
  {Onodera}, {Rizzo}, {Lee}, {Strait}, \& {Toft}}]{Valentino2023}
{Valentino}, F., {Brammer}, G., {Gould}, K. M.~L., {et~al.} 2023, arXiv
  e-prints, arXiv:2302.10936, \dodoi{10.48550/arXiv.2302.10936}

\bibitem[{Wainer \& Thissen(1976)}]{Wainer1976}
Wainer, H., \& Thissen, D. 1976, Psychometrika, 41, 9–34,
  \dodoi{10.1007/bf02291695}

\bibitem[{{Windhorst} {et~al.}(2018){Windhorst}, {Timmes}, {Wyithe},
  {Alpaslan}, {Andrews}, {Coe}, {Diego}, {Dijkstra}, {Driver}, {Kelly}, \&
  {Kim}}]{Windhorst2018}
{Windhorst}, R.~A., {Timmes}, F.~X., {Wyithe}, J. S.~B., {et~al.} 2018, \apjs,
  234, 41, \dodoi{10.3847/1538-4365/aaa760}

\bibitem[{{Windhorst} {et~al.}(2023){Windhorst}, {Cohen}, {Jansen}, {Summers},
  {Tompkins}, {Conselice}, {Driver}, {Yan}, {Coe}, {Frye}, {Grogin},
  {Koekemoer}, {Marshall}, {O'Brien}, {Pirzkal}, {Robotham}, {Ryan}, {Willmer},
  {Carleton}, {Diego}, {Keel}, {Porto}, {Redshaw}, {Scheller}, {Wilkins},
  {Willner}, {Zitrin}, {Adams}, {Austin}, {Arendt}, {Beacom}, {Bhatawdekar},
  {Bradley}, {Broadhurst}, {Cheng}, {Civano}, {Dai}, {Dole}, {D'Silva},
  {Duncan}, {Fazio}, {Ferrami}, {Ferreira}, {Finkelstein}, {Furtak}, {Gim},
  {Griffiths}, {Hammel}, {Harrington}, {Hathi}, {Holwerda}, {Honor}, {Huang},
  {Hyun}, {Im}, {Joshi}, {Kamieneski}, {Kelly}, {Larson}, {Li}, {Lim}, {Ma},
  {Maksym}, {Manzoni}, {Meena}, {Milam}, {Nonino}, {Pascale}, {Petric},
  {Pierel}, {del Carmen Polletta}, {R{\"o}ttgering}, {Rutkowski}, {Smail},
  {Straughn}, {Strolger}, {Swirbul}, {Trussler}, {Wang}, {Welch}, {B. Wyithe},
  {Yun}, {Zackrisson}, {Zhang}, \& {Zhao}}]{Windhorst2023}
{Windhorst}, R.~A., {Cohen}, S.~H., {Jansen}, R.~A., {et~al.} 2023, \aj, 165,
  13, \dodoi{10.3847/1538-3881/aca163}

\bibitem[{{Zhang} {et~al.}(2018){Zhang}, {Yu}, \& {Lu}}]{Zhang2018}
{Zhang}, C., {Yu}, Q., \& {Lu}, Y. 2018, \apj, 855, 36,
  \dodoi{10.3847/1538-4357/aaab4c}

\bibitem[{{Zitrin} {et~al.}(2013){Zitrin}, {Menanteau}, {Hughes}, {Coe},
  {Barrientos}, {Infante}, \& {Mandelbaum}}]{Zitrin2013}
{Zitrin}, A., {Menanteau}, F., {Hughes}, J.~P., {et~al.} 2013, \apjl, 770, L15,
  \dodoi{10.1088/2041-8205/770/1/L15}

\bibitem[{{Zitrin} {et~al.}(2009){Zitrin}, {Broadhurst}, {Umetsu}, {Coe},
  {Ben{\'{\i}}tez}, {Ascaso}, {Bradley}, {Ford}, {Jee}, {Medezinski},
  {Rephaeli}, \& {Zheng}}]{Zitrin2009}
{Zitrin}, A., {Broadhurst}, T., {Umetsu}, K., {et~al.} 2009, \mnras, 396, 1985,
  \dodoi{10.1111/j.1365-2966.2009.14899.x}

\bibitem[{{Zitrin} {et~al.}(2015){Zitrin}, {Labb{\'e}}, {Belli}, {Bouwens},
  {Ellis}, {Roberts-Borsani}, {Stark}, {Oesch}, \& {Smit}}]{Zitrin2015}
{Zitrin}, A., {Labb{\'e}}, I., {Belli}, S., {et~al.} 2015, \apjl, 810, L12,
  \dodoi{10.1088/2041-8205/810/1/L12}

\end{thebibliography}

\appendix
The arc systems in the El Gordo Field which are utilized in our lens model are presented  in Table~\ref{tab_2}.  The image system designations for numbers 1--23 follow \citet{Caminha2022}, and the image system designations for numbers 24--60 follow \citet{Diego2023}. We  removed {six} of the 60 previously known systems for the following reasons: systems 31 and 44 and image 43c are at or below the NIRCam detection limit, system 34 has counterimages which have different colors and one image that is very close in projection to another source, preventing its verification, and systems 45 and 50 are very close in projection to a cluster member, preventing their verifications.  At the same time, systems 31 and 43a,b have image family members which are too closely-separated to usefully constrain the macro lens model that is the objective of this study. In general, an arc system that is predominantly lensed by a single cluster member is left out. 
The one exception is for El Anzuelo (system 24) because it is situated relatively close to the cluster center and has a new tentative spectroscopic redshift presented in this paper.  The table columns are:  ID, Right Ascension, Declination, observed AB magnitude ({\sc SExtractor mag\_auto}) in the {F200W} filter, spectroscopic redshift ($z_{sp}$), photometric redshift estimate ($z_{ph}$), lens-model predicted redshift ($z_{mod}$), magnification factor ($\mu$), and the discovery citation. Among the known image systems 1 through 60, we identified an additional five counterimages (labeled  ``This study" in the `Ref.' column). We also extended the catalog to include the two new image systems 61 and 62, which meet our selection criteria. The analysis of the strong lensing properties appears in \S\ref{sec:LTM}. 

The galaxies with photometric redshifts that place them in the galaxy overdense region at $z =4.32$ are presented in Table~\ref{tab_3}. The columns are: ID, Right Ascension, Declination, observed AB magnitude ({\sc SExtractor mag\_auto}) in the {F200W} filter, photometric redshift ($z_{ph})$, and the magnification factor on the assumption that the redshift is $z=4.32$. One image from the doubly-imaged system 28 also has a photometric redshift and a model-predicted redshift of $4.40$. These identifications are all new to this study and are displayed also in Figure~\ref{fig_z4p3}. \S\ref{sec:odense} gives details on their selection.

\begin{longtable*}[c]{c c c c c c c c c}
\caption{Strong Lensing Image Systems} \label{tab:long} \\
\hline \hline
\multicolumn{1}{c }{\textbf{ID}} & \multicolumn{1}{c}{\textbf{R.~A.}} & \multicolumn{1}{c}{\textbf{Decl.}} &
\multicolumn{1}{c}{\textbf{$m_{\rm F200W,obs}$}} &
\multicolumn{1}{c}{\textbf{$z_{\rm sp}$}} &
\multicolumn{1}{c}{\textbf{$z_{\rm ph}$}} &
\multicolumn{1}{c}{\textbf{$z_{\rm mod}$}} &
\multicolumn{1}{c}{\textbf{$\mu$}} &
\multicolumn{1}{c}{\textbf{Ref.}} \\ \hline 
\endfirsthead
\multicolumn{3}{c}%
{{\bfseries \tablename\ \thetable{} -- continued from previous page}} \\
\hline \hline
\multicolumn{1}{c}{\textbf{ID}} & \multicolumn{1}{c}{\textbf{R.~A.}} & \multicolumn{1}{c}{\textbf{Decl.}} &
\multicolumn{1}{c}{\textbf{$m_{\rm F200W,obs}$}} &
\multicolumn{1}{c}{\textbf{$z_{\rm sp}$}} &
\multicolumn{1}{c}{\textbf{$z_{\rm ph}$}} &
\multicolumn{1}{c}{\textbf{$z_{\rm mod}$}} &
\multicolumn{1}{c}{\textbf{$\mu$}} &
\multicolumn{1}{c}{\textbf{Ref.}} \\ 
\hline
\endhead
\multicolumn{8}{r}{{Continued on next page}} \\
\endfoot
\hline \hline
\endlastfoot
%
1.1a&01:02:55.39&$-$49:15:00.35&22.48 &2.5636$^a$&&\no& 2.4 &Z13(1) \\ 
1.1b&01:02:53.35&$-$49:15:16.32&21.58&2.5636$^a$&&\no& 230 &Z13(1) \\ 
1.1c&01:02:52.77&$-$49:15:18.68&20.38&2.5636$^a$&$1.04^{+0.04}_{-0.07}$&\no&3.5&Z13(1) \\
1.2a&01:02:55.33&$-$49:15:01.16&24.70&2.5636$^b$&&\no&2.5  &Z13(1) \\
1.2b&01:02:53.35&$-$49:15:16.32&\no  &2.5636$^b$&&\no& 28 &Z13(1) \\
1.2c&01:02:52.61&$-$49:15:19.69&\no  &2.5636$^b$&&\no& 1.8  &Z13(1) \\
\hline
2.1a&01:02:56.58&$-$49:15:47.09&22.66&2.8254$^a$& &\no & 65 &Z13(2) \\
2.1b&01:02:55.98&$-$49:15:51.23&22.70 &2.8254$^a$& &\no & 19 &Z13(2)  \\
2.1c&01:02:54.38&$-$49:16:04.53&23.74 &2.8254$^a$& &\no & 3.3 &Z13(2)  \\
2.2a&01:02:57.09&$-$49:15:43.54&23.49&2.8254$^b$& &\no & 12 &Z13(2)  \\
2.2b&01:02:55.62&$-$49:15:53.82&23.41 &2.8254$^b$& $3.40^{+0.08}_{-0.05}$ &\no & 13 &Z13(2)  \\
2.2c&01:02:54.45&$-$49:16:04.07&23.69 &2.8254$^b$& $2.65^{+0.02}_{-0.01}$ &\no & 3.5&Z13(2)  \\
2.5a&01:02:56.52&$-$49:15:47.51&\no &2.8254$^b$& &\no & 130&Z13(2)  \\
2.5b&01:02:56.08&$-$49:15:50.57&23.53 &2.8254$^b$& &\no &23&Z13(2)  \\
\hline
3a  &01:02:51.64&$-$49:14:53.67&$28.19$&3.3300$^a$& &\no& 18 &Cam22(3a) \\ 
3b  &01:02:50.69&$-$49:15:06.54&$28.68$&3.3300$^a$& & \no  & 14 &Cam22(3b) \\ 
\hline
4b &01:02:52.41&$-$49:15:03.10&25.98& 3.3339$^a$& &\no & 20 &Cam22(4b) \\ 
4c &01:02:51.65&$-$49:15:09.28&25.27& 3.3339$^a$& &\no & 4.6 &Cam22(4c) \\ 
\hline
5.1a&01:02:59.99&$-$49:15:49.47&24.09&3.5376$^a$ & $4.01^{+0.02}_{-0.02}$&\no& 6.8 &Z13(4.1) \\
5.2a&01:02:59.99&$-$49:15:50.64&25.68 &3.5376$^b$ & $3.85^{+0.03}_{-0.06}$ &\no& 8.2 & \\
5.3a&01:02:59.97&$-$49:15:50.30&26.41&3.5376$^b$ & $3.22^{+0.15}_{-0.13}$ &\no& 8.8 &\\
5.1b&01:02:56.62&$-$49:16:08.21&23.90&3.5376$^a$ &$4.01^{+0.04}_{-0.06}$ & \no& 63 &Z13(4.5) \\
5.2b&01:02:56.74&$-$49:16:08.38&25.49&3.5376$^b$ & $3.90^{+0.03}_{-0.14}$ & \no& 20 &  \\
5.3b&01:02:56.71&$-$49:16:07.93&25.05 &3.5376$^b$ & $3.90^{+0.02}_{-0.05}$ & \no& 28 & \\
5.1c&01:02:55.37&$-$49:16:26.03&23.94&3.5376$^a$ & $4.04^{+0.01}_{-0.01}$ & \no&2.2&Z13(4.4) \\
5.2c&01:02:55.39&$-$49:16:27.23&25.41&3.5376$^b$ & $3.92^{+0.04}_{-0.05}$ & \no&2.1&  \\
5.3c&01:02:55.34&$-$49:16:26.87&26.22&3.5376$^b$ & $3.85^{+0.03}_{-0.06}$ & \no&2.1&  \\
\hline
6a&01:02:59.45&$-$49:15:54.71&27.86&4.1879$^a$& $4.04^{+0.18}_{-0.09}$&\no& 4.7 & Cam22(6a)\\
6b&01:02:57.80&$-$49:16:03.32&$27.78$&4.1879$^a$& &\no& 6.8 & Cam22(6b)\\
6c&01:02:54.45&$-$49:16:31.76&$28.10$&4.1879$^c$& &\no& 1.8 & D23\\
\hline
7a&01:02:57.59&$-$49:15:15.55&26.43&4.2306$^c$& $3.74^{+0.05}_{-0.04}$ &\no& 3.6 & D23(7a)\\
7b&01:02:54.59&$-$49:15:37.22&25.69&4.2306$^a$& $0.51^{+4.00}_{-0.23}$$^d$ &\no& 5.4 & Cam22(7b)\\
7c&01:02:53.03&$-$49:15:48.74&26.10&4.2306$^a$& $4.09^{+0.13}_{-0.20}$  &\no& 5.8 & Cam22(7c)\\
\hline
8a&01:02:55.99&$-$49:15:05.38&25.29&4.3175$^e$& $4.12^{+0.08}_{-0.09}$ &\no& 4.2 & Ca21(2a) \\
8b&01:02:54.61&$-$49:15:16.44&\no& 4.3175$^c$& & {\no}& 3.3 & Ca21(2b)\\
8c&01:02:51.24&$-$49:15:37.08&25.93& 4.3175$^c$& $4.01^{+0.26}_{-0.15}$&\no& 3.1 & D20; Ca21(2c) \\
\hline
9.1a&01:02:55.78&$-$49:15:07.89&22.20 &4.3196$^e$& $4.47^{+0.02}_{-0.02}$\rlap{$^d$}&\no&5.6&Ca21(4a) \\
9.1b&01:02:54.85&$-$49:15:14.90&\no&4.3196$^c$ & &{\no}&4.0&Ca21(4b) \\
9.1c&01:02:51.11&$-$49:15:38.53&24.21&4.3196$^c$ & $4.29^{+0.09}_{-0.08}$&\no&3.0 &Ca21(4c) \\
9.2a&01:02:55.74&$-$49:15:08.37&\no&4.3196$^b$& &\no& 7.5 &D23(9.2a) \\
9.2b&01:02:54.89&$-$49:15:14.61&\no&4.3196$^b$& &\no& 4.2 &D23(9.2b) \\
9.2c&01:02:51.06&$-$49:15:38.86&\no&4.3196$^b$& &\no& 2.9 &D23(9.2c)  \\
\hline
10a&01:02:56.28&$-$49:15:06.93&24.07 & 4.3275$^e$ & $4.42^{+0.07}_{-0.05}$ &\no& 4.0 & Z13(3.1),Ca21(3a)\\
10b&01:02:54.77&$-$49:15:19.53&\no& 4.3269$^e$ &&\no& 3.4 & Z13(3.2),Ca21(3b)\\
10c&01:02:51.56&$-$49:15:38.36&24.71& 4.3289$^e$ &$4.29^{+0.06}_{-0.08}$&\no& 3.0 & Z13(3.3),Ca21(3c)\\
\hline
11.1a&01:02:55.82&$-$49:15:00.30&23.58&4.3278$^e$ & $4.27^{+0.07}_{-0.06}$ &\no & 4.5& D20;Ca21(1a) \\
11.2a&01:02:55.86&$-$49:14:59.37&24.26 &4.3278$^b$ &$4.29^{+0.06}_{-0.05}$ & \no& 4.2 &  \\
11.1b&01:02:54.31&$-$49:15:12.43&\no &4.3273$^e$ & & \no& 3.0 & D20;Ca21(1b)\\
11.2b&01:02:54.27&$-$49:15:12.33&\no &4.3273$^b$ & &\no &2.9&  \\
11.1c&01:02:50.97&$-$49:15:33.55&24.52 &4.3273$^e$ &$4.34^{+0.09}_{-0.07}$ &\no &3.2& D20;Ca21(1c) \\
11.2c&01:02:50.97&$-$49:15:33.08&\no &4.3273$^b$ & &\no &3.0&  \\
\hline
12.1a&01:02:55.44&$-$49:14:49.23&25.39&4.7042$^c$& &\no& 2.5 &Cam22(12a)\\
12.2a&01:02:55.44&$-$49:14:49.83&25.94&4.7042$^c$& $4.88^{+0.01}_{-0.01}$ &\no&2.6 &Cam22(12a)\\
12.1b&01:02:53.38&$-$49:15:04.29&26.87&4.7042$^a$& &\no& 7.0 &Cam22(12b)\\
12.2b&01:02:53.43&$-$49:15:04.95&\no&4.7042$^a$& &\no&7.2&Cam22(12b)\\
12.1c&01:02:50.43&$-$49:15:26.15&25.20&4.7042$^a$& $4.70^{+0.16}_{-0.09}$&\no& 5.1 &Cam22(12c)\\
12.2c&01:02:50.43&$-$49:15:26.69&\no&4.7042$^a$& &\no&4.8&Cam22(12c)\\
\hline
13a&01:02:56.86&$-$49:15:00.06&$28.38$&4.7528$^c$& &\no& 2.5 &D23 \\ 
13b&01:02:54.41&$-$49:15:26.49&$28.33$&4.7528$^a$& &\no& 4.8 &Cam22(13b) \\ 
13c&01:02:52.19&$-$49:15:39.44&$>$27.93&4.7528$^a$& &\no& 6.2 &Cam22(13c) \\ 
\hline
14a&01:02:57.51&$-$49:15:24.17&$>$27.93&4.9486$^a$& &\no& 12 &Cam22(14a) \\
14b&01:02:55.73&$-$49:15:35.60&$>$27.93&4.9486$^a$& &\no& 6.3 &Cam22(14b) \\
\hline
15b&01:02:55.13&$-$49:16:09.01&24.72 &4.9770$^a$& $4.95^{+0.03}_{-0.02}$$^d$ &\no& 5.5 &Cam22(15b) \\ 
15c&01:02:54.98&$-$49:16:10.74&24.72 &4.9770$^a$& &\no&5.0&Cam22(15c) \\
\hline
16a&01:02:56.15&$-$49:15:16.70&$>$27.93&5.0880$^a$&&\no & 11 & Cam22(16a)\\ 
16b&01:02:55.56&$-$49:15:21.71&$>$27.93&5.0880$^a$& &\no& 32 & Cam22(16b)\\ 
16c&01:02:51.57&$-$49:15:49.93&$>$27.93&5.0880$^c$& &\no& 2.7 & This paper\\ 
\hline
17a&01:02:50.29&$-$49:14:53.04&$27.89$&5.0929$^a$& &$-$& 13 & Cam22(17a)\\ 
17b&01:02:50.63&$-$49:14:55.36&$26.75$&5.0929$^a$& & \no& 12 & Cam22(17b)\\ 
17c&01:02:50.36&$-$49:14:58.58&26.64&5.0929$^a$& $5.15^{+0.03}_{-0.03}$$^d$ &\no& 4.0 & Cam22(17c)\\ 
\hline
18a&01:02:57.66&$-$49:15:11.80&$>$27.93  &5.1173$^c$&& \no & 3.2 & D23\\ 
18b&01:02:54.67&$-$49:15:33.62&$>$27.93  &5.1173$^a$&& \no & 3.8 & Cam22(18b)\\ 
18c&01:02:52.57&$-$49:15:49.96&26.65&5.1173$^a$& $5.16^{+0.06}_{-0.10}$ & \no& 4.3 & Cam22(18c)\\ 
\hline
19a&01:02:58.18&$-$49:15:11.16&26.67&5.1199$^c$& $5.41^{+0.06}_{-0.11}$&\no& 2.7 & D23\\ 
19b&01:02:53.80&$-$49:15:40.74&27.12&5.1196$^a$&$1.89^{+1.73}_{-0.25}$ &\no& 5.9 & Cam22(19b)\\
19c&01:02:53.23&$-$49:15:45.56&26.86&5.1196$^a$& &\no& 16 & Cam22(19c)\\ 
\hline
20a&01:03:00.70&$-$49:15:43.22&27.04&5.4845$^c$& &\no& 6.7 & D23\\
20b&01:02:56.07&$-$49:16:15.15&26.95&5.4845$^a$&$5.85^{+0.28}_{-0.15}$ &\no& 3.0 & Cam22(20b)\\
20c&01:02:55.67&$-$49:16:22.06&27.74&5.4845$^a$& $4.49^{+0.18}_{-0.29}$&\no& 3.0 & Cam22(20c)\\
\hline
21a&01:02:57.80&$-$49:15:24.98&$>$27.93&5.5811$^a$& &\no& 11 & Cam22(21a) \\ 
21b&01:02:55.98&$-$49:15:37.76&$>$27.93&5.5811$^a$& &\no& 4.3 & Cam22(21b) \\ 
\hline
22.1a&01:03:01.39&$-$49:16:15.23&26.96&5.9521$^a$&&\no & 5.2 & Cam22(22a)\\
22.2a&01:03:01.28&$-$49:16:16.76&27.39&5.9521$^c$&$6.00^{+0.48}_{-0.15}$&\no &2.0& D23\\
22.1b&01:03:00.14&$-$49:16:35.72&26.72&5.9521$^a$& $1.12^{+0.03}_{-0.03}$ &\no& 2.0 & Cam22(22b)\\
22.2b&01:03:00.24&$-$49:16:34.40&$28.31$&5.9521$^c$&\no&6.5& D23\\
\hline
23.3a&01:02:59.64&$-$49:16:26.36&\no&2.1878$^a$& &\no& 3.0 &Z13(c5.1)\\
23.3b&01:02:59.38&$-$49:16:29.17&\no&2.1878$^a$& &\no& 2.5&Z13(c5.2)\\
23.3c&01:02:57.95&$-$49:16:38.88&24.52&2.1878$^c$&&\no & 1.8 &D23\\
\hline
24.1a&01:02:49.215&$-$49:15:08.84 &21.23&2.29$^f$&$2.00^{+0.01}_{-0.01}$& & 5.8 &D23\\
24.1b&01:02:49.322&$-$49:15:04.52 &\no&2.29$^f$&& & 6.6 &D23\\
24.1c&01:02:49.452&$-$49:15:06.03 &\no&2.29$^f$&& &6.1&D23\\
24.2a&01:02:49.150&$-$49:15:05.44 &21.90&2.29$^f$& $2.70^{+0.30}_{-0.02}$&&5.5&D23\\
24.2b&01:02:49.141&$-$49:15:05.99 &\no&2.29$^f$ & && 5.6&D23\\
24.2c&01:02:49.263&$-$49:15:03.86 &\no&2.29$^f$ && &6.2&D23\\
\hline
25a&01:02:54.55&$-$49:14:58.59&22.40&  & & 2.44 &2.5 &D23\\
25b&01:02:53.26&$-$49:15:06.85&21.62&  & &      & 4.3&D23\\
25c&01:02:51.83&$-$49:15:16.91&23.01&  & &      &4.4 &D23\\
\hline
26a&01:03:00.26&$-$49:15:52.35&$27.77$& &  & 3.34 &9.7&D23  \\ 
26b&01:02:56.89&$-$49:16:12.27&28.99& & & &12&D23  \\
26c&01:02:55.88&$-$49:16:28.44&27.81& & $0.11^{+0.07}_{-0.05}$& &&D23  \\
\hline
27a&01:02:56.33&$-$49:16:16.09&$>$27.93& & & 3.82 &10 &D23  \\
27b&01:02:56.24&$-$49:16:17.35&$>$27.93& & &      &7.7&D23 \\
\hline 
28a&01:02:55.51&$-$49:16:07.15&25.67& &$4.34^{+0.12}_{-0.12}$& 4.40& 6.6 &Z13(c7.1)\\
28b&01:02:54.94&$-$49:16:14.73&25.38& &$3.24^{+0.12}_{-0.20}$ &&6.5& Z13(c7.2)\\
\hline
29a&01:02:55.85&$-$49:16:07.45&24.67&&  & 3.28 & 8.0 & D23\\
29b&01:02:55.24&$-$49:16:15.79&25.24&& $3.42^{+0.02}_{-0.03}$&&3.0& D23\\
29c&01:02:59.72&$-$49:15:43.77&25.93&&$3.69^{+0.05}_{-0.06}$ &&4.6& D23\\
\hline
30a&01:02:56.32&$-$49:16:07.80&24.54&&$2.89^{+0.08}_{-0.08}$ & 2.25 & 6.0 & Z13(c9.1)\\
30b&01:02:55.66&$-$49:16:17.43&24.21&&$2.25^{+0.08}_{-0.10}$& &2.4& Z13(c9.2) \\
30c&01:02:59.07&$-$49:15:53.23&24.57&&$2.46^{+0.23}_{-0.07}$& & 3.4& Z13(c9.3)\\
\hline
32a&01:02:54.59&$-$49:14:54.12&24.91& \no & $3.77^{+0.02}_{-0.01}$& 2.50 & 2.6 & D23\\ 
32b&01:02:53.04&$-$49:15:04.85&25.39& \no && & 5.1 & D23\\ 
32c&01:02:51.80&$-$49:15:14.24&25.19 & \no &$3.75^{+0.01}_{-0.02}$& & 3.1 & D23\\ 
\hline
33a&01:02:59.89&$-$49:16:30.64&26.03& &$2.67^{+0.07}_{-0.09}$  & 10.87 &44&D23 \\
33b&01:02:59.74&$-$49:16:32.47&26.05& &$3.84^{+0.13}_{-0.05}$&& 11 &D23 \\
\hline
35a&01:02:58.53&$-$49:16:36.96&25.63& &$1.06^{+0.03}_{-0.06}$& 2.23 &3.1&D23 \\
35b&01:02:58.75&$-$49:16:35.70&25.94& &$0.20^{+0.01}_{-0.01}$ &&3.3&D23 \\
35c&01:03:00.13&$-$49:16:20.82&25.46& &$2.64^{+0.55}_{-0.03}$ &&4.6&D23 \\
\hline
36a&01:02:57.32&$-$49:15:45.39&24.95& &$3.45^{+0.01}_{-0.00}$  & 2.96 &15&D23 \\
36b&01:02:55.79&$-$49:15:56.19&\no& &&&10&D23 \\
36c&01:02:54.60&$-$49:16:06.90&$27.22$& &&&3.0&D23 \\
\hline
37a&01:02:55.76&$-$49:16:08.94&26.47& &$0.69^{+0.06}_{-0.13}$& 9.86 &2.9&D23 \\
37b&01:02:55.07&$-$49:16:19.71&$>$27.93& & &&12&D23 \\
\hline
38a&01:02:58.97&$-$49:15:55.90&23.97&&$1.68^{+0.00}_{-0.01}$ &3.08&1.5&D23 \\
38b&01:02:58.72&$-$49:15:56.85&24.37& &$3.77^{+0.02}_{-0.01}$&&2.0&D23 \\
38c&01:02:58.67&$-$49:15:57.29&24.85& &$3.71^{+0.01}_{-0.01}$&&2.2&D23 \\
38e&01:02:54.76&$-$49:16:27.93&25.46& &$3.40^{+0.03}_{-0.03}$&&1.9&D23 \\
\hline
39a&01:02:55.03&$-$49:15:03.00&25.83& &$4.29^{+0.16}_{-0.16}$ & 4.14 &3.8&D23 \\ 
39b&01:02:54.42&$-$49:15:08.30&24.09& && &7.2 &D23 \\ 
39c&01:02:50.10&$-$49:15:28.77&$26.14$& && & 3.3&This Paper \\
\hline
40a&01:02:56.63&$-$49:15:47.43&$27.72$ & & & 3.03 & 23 & D23 \\
40b&01:02:55.96&$-$49:15:52.05&$27.52$ & & && 28& D23\\
40c&01:02:54.51&$-$49:16:03.71&$26.67$ &  & &&3.8& D23 \\
\hline
41a&01:02:51.53&$-$49:15:05.90&25.99& & $5.65^{+0.09}_{-0.18}$& 4.02& 6.0 &D23\\
41b&01:02:51.66&$-$49:15:04.68&25.47& && & 6.9 &D23 \\
\hline
42a&01:02:57.98&$-$49:15:59.14&25.16&&$2.59^{+0.19}_{-0.13}$ & 2.24& 11&D23\\  
42b&01:02:57.64&$-$49:15:59.80&$27.75$&& & &14&D23 \\
42c&01:02:55.52&$-$49:16:19.98&25.99& &$3.76^{+0.06}_{-0.07}$& & 2.2&D23 \\
\hline 
46a&01:02:58.759&$-$49:16:45.35&$23.68$ & & &  2.76 &2.6& D23 \\
46b&01:02:58.499&$-$49:16:46.27&$23.68$ & && &2.5& D23 \\
\hline
47a&01:02:51.25&$-$49:15:01.35&\no  & & & 3.71 &8.9&D23 \\
47b&01:02:51.08&$-$49:15:03.35&28.74& & && 9.6&D23 \\ 
\hline
48a&01:02:54.62&$-$49:14:45.72&25.88& & $3.06^{+0.02}_{-0.06}$& 4.31 &2.6&D23 \\ 
48b&01:02:50.92&$-$49:15:13.71&25.96& &$0.98^{+0.18}_{-0.15}$ &&7.8&D23 \\
\hline
49a&01:02:53.27&$-$49:15:14.55&28.52& & & 2.46 &39&D23 \\
49b&01:02:52.78&$-$49:15:16.31&$27.18$& & &&17&D23 \\
49c&01:02:55.19&$-$49:14:59.66&24.85& &$3.81^{+0.03}_{-0.05}$ && 2.6&D23 \\ 
\hline
51a&01:02:53.47&$-$49:15:14.26&\no     && & 2.77 &16&D23\\
51b&01:02:52.46&$-$49:15:18.40&$25.61$ && &      &12&D23 \\
51c&01:02:55.22&$-$49:14:59.35&$25.18$ && &      &2.5&D23 \\
\hline
52a&01:02:55.86&$-$49:15:51.85&$26.07$& & & 2.92 & 34 &D23\\ %
52b&01:02:56.74&$-$49:15:45.66&$26.12$ && & & 25 &D23\\ %
52c&01:02:54.42&$-$49:16:03.93&\no & && & 3.9 &D23\\ 
\hline
53a&01:02:58.52&$-$49:15:39.30&26.48& &&7.11 &7.4 & D23\\ 
53b&01:02:59.13&$-$49:15:40.12&$27.82$& & && 58& D23 \\
53c&01:02:58.77&$-$49:15:38.83&$27.33$& & && 19& D23\\
\hline
54a&01:02:54.24&$-$49:16:05.14&26.30 & &$4.08^{+0.14}_{-0.15}$ & 4.30 & 6.7 &D23\\
54b&01:02:55.01&$-$49:15:58.00&26.13 & &$3.70^{+0.10}_{-0.09}$ && 6.1 &D23 \\ 
54c&01:02:59.08&$-$49:15:31.23&26.83 & &$3.65^{+0.02}_{-0.03}$ & &5.4&D23 \\
\hline
55a&01:02:52.74&$-$49:15:21.11&26.37& &$1.25^{+0.14}_{-0.16}$ & 3.18 &1.5&D23\\ 
55b&01:02:52.81&$-$49:15:20.85&25.53&&$4.41^{+0.25}_{-0.09}$ & & 1.6&D23 \\ 
55c&01:02:55.58&$-$49:14:56.94&24.19&&$2.09^{+0.17}_{-0.10}$ & & 1.1&This paper \\ 
\hline
56a&01:02:59.11&$-$49:15:54.18&22.44&&$3.18^{+0.02}_{-0.02}$ & 3.20 &2.3 &D23\\
56b&01:02:57.56&$-$49:15:59.80&22.31& && & 5.9&D23 \\
56c&01:02:54.78&$-$49:16:27.06&23.43& &$3.32^{+0.09}_{-0.08}$& & 1.9&D23\\
56d&01:02:57.69&$-$49:16:02.07&\no& & && 13&D23 \\
\hline
57a&01:02:59.63&$-$49:15:56.55&$>$27.93& & & 4.15 &28&D23\\ 
57b&01:02:58.45&$-$49:16:03.37&27.80& &$3.21^{+0.19}_{-0.07}$& & 3.1&D23\\ 
\hline 
58a&01:02:58.91&$-$49:16:42.87&$28.05$& & & 3.40 & 2.3 &D23\\
58b&01:02:59.02&$-$49:16:42.21&$>$27.93& & && 2.2 &D23 \\
\hline 
59a&01:02:52.67&$-$49:15:21.17&$>$27.93& & & 4.57 &2.0&D23\\
59b&01:02:52.76&$-$49:15:20.69&27.44& & && 1.7 &D23 \\
59c&01:02:56.43&$-$49:14:51.46&27.45& &$0.75^{+0.11}_{-0.16}$ & & 2.2 &This paper\\
\hline
60a&01:02:53.83&$-$49:15:17.40&26.53& & $1.04^{+0.05}_{-0.04}$& 2.66 &12&D23\\
60b&01:02:52.70&$-$49:15:23.04&26.59& & $1.06^{+0.34}_{-0.14}$ & &3.4&D23\\
60c&01:02:55.31&$-$49:15:05.89&26.50& &$1.70^{+0.35}_{-0.37}$ & &2.2&This paper\\
\hline
61a&01:02:56.54&$-$49:15:25.61&$>$27.93& & &3.78 &21&This paper\\
61b&01:02:55.64&$-$49:15:32.06&$>$27.93& & & &20&This paper\\
\hline
62a&01:02:55.83 & -49:16:03.35 & $26.05$  & 4.863$^f$  & & & 10& This paper\\
62b&01:02:54.60 &-49:16:19.91 & $27.90$& 4.863$^f$ & & &  2.9& This paper\\
62c&01:03:00.98 &-49:15:38.32 &$25.72$ & 4.863$^f$ & & & 2.4& This paper 
\label{tab_2}
\end{longtable*}
\tablenotetext{a}{A spectroscopic redshift is measured at this coordinate in \citet{Caminha2022}.}
\tablenotetext{b}{A spectroscopic redshift is assumed along the same arc and near to this specific coordinate in projection.}
\tablenotetext{c}{A spectroscopic redshift is measured not at this coordinate, but for one or more of its counterimages.}
\tablenotetext{d}{The photometric redshift is estimated using EAZY.}
\tablenotetext{e}{A spectroscopic redshift is measured at this coordinate in \citet{Caputi2021}.}
\tablenotetext{f}{A spectroscopic redshift is measured for this arc in this study.}

 \clearpage
 
\begin{deluxetable}{cccccc}
\tablecaption{New $z=4.32$ Galaxy Candidates}
\tablecolumns{6}
\tablehead{
 \colhead{{\bf ID}} & \colhead{{\bf R.~A.}} &  \colhead{{\bf Decl.}} & \colhead{$m_{\rm F200W,obs}$} & \colhead{$z$} & \colhead{$\mu$}
}
\startdata
\multicolumn{6}{l}{Found with MUSE single-line detection:}\\
101 & 01:02:51.20 & $-$49:15:27.01 &\llap{$>$}27.96&4.317& 2.1  \\
\hline
\multicolumn{6}{l}{Found with LePhare photometric redshift estimates:}\\
102 &01:02:48.57&$-$49:16:00.26&26.34&$4.12^{+0.24}_{-0.13}$&1.5 \\ 
103 &01:02:50.46&$-$49:16:05.99&27.29&$4.22^{+0.26}_{-3.4} $&1.7 \\ 
104 &01:02:50.46&$-$49:16:21.07&27.81&$3.54^{+0.86}_{-0.09} $& 1.6 \\ 
105 &01:02:51.16&$-$49:16:18.84&24.74&$4.20^{+0.11}_{-0.09} $&1.7 \\ 
106 &01:02:52.91&$-$49:16:09.12&28.02&$4.20^{+0.64}_{-0.64} $&2.5 \\ 
28a &01:02:55.51&$-$49:16:07.15&25.67&$4.34^{+0.12}_{-0.12} $&6.6\\ 
107 &01:02:55.83&$-$49:15:49.25&26.31&$5.65^{+0.21}_{-4.7} $ &11\phantom{.00} \\ 
108 &01:02:57.78&$-$49:16:42.92&25.55&$4.20^{+0.13}_{-0.14} $&2.2 \\ 
109 &01:02:57.97&$-$49:16:44.33&27.57&$3.97^{+0.52}_{-0.38} $&2.2 \\ 
110&01:02:58.24&$-$49:15:19.04&28.00&$4.85^{+0.05}_{-1.4} $ & 3.6 \\ 
111&01:03:01.39&$-$49:14:40.88&24.81&$4.22^{+0.09}_{-0.14} $&1.6 \\ 
112&01:03:01.83&$-$49:15:40.50&26.10&$4.60^{+0.21}_{-3.7} $ & 2.2 \\
  \label{tab_3}
\enddata
\tablecomments{Column 4 gives measured AB magnitudes, not corrected for magnification.}
\end{deluxetable}
\end{document}